# Adaptive shaping for generation of optical arbitrary waveform

Fangming Liu[*]

A new adaptive shaping method that can generate arbitrary optical waveforms with folded-type or fan-type birefringent variable shapers is proposed in this paper. Automatic arbitrary laser temporal shaping of picosecond and femtosecond lasers in shapers with material dispersion, nonidentical birefringent retarders are also studied and results are presented demonstrating the power of the method.

## 1. Introduction

Laser temporal shaping for producing pulses with arbitrary temporal profiles is important for many applications. Recently, laser temporal shaping with folded-type and fan-type birefringent variable shapers were experimentally demonstrated for generating various shaped pulses with high fidelity [1, 2]. The structure of the folded-type and fan-type shapers are based on the Solc folded and fan filters concepts [3]. Different from the conventional variable shapers like Dazzler [4], grating based 4f spectral shaping systems [5] and nanosecond pulse shaping systems [6], which require the laser pulse have relatively wide bandwidth or duration, the folded-type and fan-type birefringent variable shapers are also applicable to narrowband picosecond pulse shaping. In many situations, the complete knowledge for pulse shaping (e.g. input laser pulse, the target output pulse and shaper's characteristics) is generally unknown exactly beforehand, resulting the desired shaper's parameters can't be directly computed. Therefore, self-learning shaping method is needed for those situations to adaptively shaping the lasers to generate the desired pulses. Besides, when the number of birefringent retarders (BRs) in folded-type or fan-type shaper is not too much, it is feasible to adjust the BRs manually to generate the laser pulse with simple target profiles, like flattop, triangular, sawtooth, etc. However, for shapers containing large number of BRs (e.g. a number of dozens), which is required for generating pulse with very complicated target profile, it would be very difficult or even unpractical to generate high fidelity shaped pulse by adjusting the shaper manually.

In this paper, I propose a new adaptive shaping method that can automatically generate almost arbitrary optical waveforms with high fidelity. To my knowledge, this is the first time to report the research about the automatic arbitrary laser temporal shaping in folded-type and fan-type shapers with the material dispersion and nonidentical BRs taken into account. The method discretizes the target optical profile with $N + 1$ points, making the target optical profile be able to be connected with the elements of birefringent shaper containing $N$ number of BRs. Predefined shape of the target pulse or its induced process (e.g. second harmonic pulse in nonlinear crystal) can be used as feedback signal for the shaping algorithm to directly optimize the shaper to perform adaptive shaping. The method is applicable to the folded-type or fan-type birefringent variable shaper over a wide range of laser parameters.

## 2. Pulse shaping with nonidentical birefringent retarders

The folded-type and fan-type birefringent variable shapers are composed by two polarizers and a number of BRs, which are designed to bring the equal time delay between the ordinary (o) and extraordinary (e) light pulses [1, 2]. A birefringent shaper containing $N$ BRs (e.g., a-cut $YVO_4$ crystals, BBO crystals) can split an input light pulse into $2^N$ replicas

Fangming Liu
[*]liu_fangming@pku.edu.cn
Institute of Heavy Ion Physics
School of Physics
Peking University
Beijing 100871, China

pulses. As each birefringent retarder (BR) brings the same time delay between the o and e replica pulses, therefore the $2^N$ replica pulses can be grouped into $N + 1$ groups according to the time delays and be transformed into $N + 1$ replica pulses. The pulse outputted from the shaper can be regarded as the coherent combination of those $N + 1$ replica pulses. According to the calculation, the folded-type and fan-type shaper have four different configurations, as shown in the Eq. (1) and Eq. (2), respectively.

$$\Theta_n = B_1(-1)^n \frac{45°}{N} + B_2, \qquad \Theta_p = 90° \qquad (1)$$

$$\Theta_n = B_1 \frac{45°}{N}(2n - 1) + B_2, \qquad \Theta_p = 0° \qquad (2)$$

Where $N$ is the total number of BRs in the shaper, and $\Theta_n$ is the angle between the slow axis of the $n - th$ BR and the polarization direction of the input polarizer, $n = 1, 2, \cdots, N$. $B_1$ and $B_2$ are constants, which are $B_1 = 1, -1$ and $B_2 = 0°, 90°$. $\Theta_p$ is the angle between the polarization directions of the output polarizer and input polarizer of the shaper. Eq. (1) and Eq. (2) depicts the crystal and polarizer arrangements in the folded-type and fan-type shapers, respectively.

Due to different physical structures, the rotation angle tuning rules of the BRs in the folded-type shaper are very different from those in fan-type shaper. According to the calculation, it is found that there are very interesting rules for tuning the rotation angles of the BRs in the shaper. The effect of increasing $\Theta_n$ on the amplitudes of all $N + 1$ replica pulses for folded-type shaper and fan-type shaper with four different configurations are presented in Table 1 and Table 2, respectively. When decreasing $\Theta_n$, the amplitude variations of the replicas are the opposite of those listed in Table 1 and Table 2. For folded-type shaper, the effect of $\Theta_n$ variation on the replicas is related to the parity of the BR sequence number $n$. However, for fan-type shaper, there aren't such kind of discrepancies between odd BR sequence number $n$ and even BR sequence number $n$. The intensity of different regions in the shaped pulse are positive related to the intensity of the respective replica pulses, as shown in Fig. A1 and Fig. A2, which are in the Appendix part and show the variations of the replicas and shaped pulse when changing the rotation angle $\Theta_{1, or\ 2}$ in folded-type and fan-type shapers.

Table 1. Amplitude variations of the replica pulses for folded-type shaper with four different configurations when increasing $\Theta_n$. Symbol explanation: e.g. "$A_n$ & $A_{n+1} \downarrow (\uparrow)$" means both the $n - th$ replica pulse amplitude $A_n$ and the $(n + 1) - th$ replica pulse amplitude $A_{n+1}$ would decrease (increase) when $n$ is an odd (even) number, at the same time all other replicas pulses' amplitudes would increase (decrease). Refer to the Fig. A1 in the Appendix.

|  | $B_2 = 90°$ | $B_2 = 0°$ |
|---|---|---|
| $B_1 = 1$ | $A_n$ & $A_{n+1} \downarrow (\uparrow)$ | $A_{N+2-n}$ & $A_{N+1-n} \downarrow (\uparrow)$ |
| $B_1 = -1$ | $A_n$ & $A_{n+1} \uparrow (\downarrow)$ | $A_{N+2-n}$ & $A_{N+1-n} \uparrow (\downarrow)$ |

Table 2. Amplitude variations of the replica pulses for fan-type shaper with four different configurations when increasing $\Theta_n$. Symbol explanation: e.g. "$A_n \uparrow$ & $A_{n+1} \downarrow$" means the $n - th$ replica pulse amplitude $A_n$ would increase and the $(n + 1) - th$ replica pulse amplitude $A_{n+1}$ would decrease, at the same time all other replicas pulses' amplitudes have little changes. Refer to the Fig. A2 in the Appendix.

|  | $B_2 = 90°$ | $B_2 = 0°$ |
|---|---|---|
| $B_1 = 1$ | $A_n \uparrow$ & $A_{n+1} \downarrow$ | $A_{N+2-n} \uparrow$ & $A_{N+1-n} \downarrow$ |
| $B_1 = -1$ | $A_n \downarrow$ & $A_{n+1} \uparrow$ | $A_{N+2-n} \downarrow$ & $A_{N+1-n} \uparrow$ |

To describe the influence of the time delay $\tau$ introduced by BR between o light and e light, the relative time delay ratio $\zeta$ is defined as

$$\zeta = \frac{\tau}{\tau_0} \qquad (3)$$

Where $\tau_0$ is the pulse width (full width at half maximum, FWHM) of the input laser pulse.

The temporal profiles of the shaped pulses can be classified into three broad categories, which are smooth profile, rippled profile and micro-pulse train according the value of $\zeta$. According to the calculation, for input pulse with Gaussian temporal profile, the $\zeta$ value ranges for generating shaped pulses with smooth profile, rippled profile and micro-pulse train are about $\zeta < 1$, $\zeta \in [1, 2.8]$, and $\zeta > 2.8$, respectively. While for input pulse with Sech$^2$ temporal profile, the corresponding $\zeta$ value ranges for generating shaped pulses with smooth profile, rippled profile and micro-pulse train are about $\zeta < 0.7$, $\zeta \in [0.7, 4]$, and $\zeta > 4$, respectively. The ripple depth of rippled pulse and the interval between micro pulses of pulse train can be adjusted by changing the relative time delay ratio $\zeta$.

The rules presented in Table 1 and Table 2 are deduced for ideal shapers which have each BR introduce the same time delay $\tau$ between the o and e light pulses. Those rules present a clear description of the replicas amplitude variation when adjusting the rotation angles of the BRs. However, the intensity of the shaped pulse synthesized by those $N + 1$ replicas are not linearly related with that of replicas, since there exists interference among the replicas. When the $\zeta$ is relatively small (e.g. $\zeta < 1$ for Gaussian input pulse) that makes the shaper generating light pulse with smooth profiles, all the replicas would be invisible in the practice, since the pulse profile measurement setups (e.g. cross-correlator, streak camera) can only measure the overall temporal profile of the shaped pulse. Therefore, it is important to establish the connection between the overall profile of shaped pulse and the elements of the shaper. On the other hand, in many situations, the time delay $\tau$ introduced by each BR (e.g. a-cut YVO$_4$, BBO crystal) wouldn't be exactly equal because of the limitation of processing precision and possible inhomogeneous in the BR material. Therefore, the $N + 1$ replica pulses may not be evenly spaced in time, and the profiles of those replicas may also not the same with each other in such non-ideal shapers.

Considering the unequal time delay $\tau$ for the BRs in the shaper, the pulse shaping under non-ideal shapers is investigated. Assuming a shaper contains eight BRs with nominal time delay $\tau$ of 3ps, and the tolerance of the BR time delay is $\pm 0.15 ps$, which means there are $\pm 5\%$ random deviations for the BR time delay. Besides, let's also introduce some deviations ($\pm 5\%$) into the BRs phase delays, i.e., $\pm 0.1\pi$ tolerance. The BR phase delay is defined as $\varphi = 2\pi(c\tau/\lambda - \lfloor c\tau/\lambda \rfloor)$, where $c$ is the speed of light in vacuum, $\lambda$ is the light wavelength, $\lfloor \ \rfloor$ is floor function that rounds down to the next integer (e.g. $\lfloor 16.8 \rfloor = 16$). The deviations for the time delay $\tau$ and phase delay $\varphi$ of each BR are listed in Table 3. Figures 1 and 2 show the simulation results of shaped pulse when changing the rotation angle of the 1st or 2nd BR in folded-type shaper and fan-type shaper, respectively. The input pulse is a Gaussian laser pulse with duration of 3ps. Therefore, the relative time delay ratio $\zeta$ is 1.

Table 3. Random deviations introduced to the time delay $\tau$ and phase delay $\varphi$ of the eight BRs for folded-type shaper and fan-type shaper.

| Deviations | 1 | 2 | 3 | 4 | 5 | 6 | 7 | 8 |
|---|---|---|---|---|---|---|---|---|
| $\Delta\tau(ps)$ | -0.067 | -0.136 | -0.121 | 0.097 | 0.058 | -0.055 | 0.135 | -0.140 |
| $\Delta\varphi(\pi * rad)$ | 0.078 | 0.092 | 0.009 | -0.072 | -0.070 | -0.048 | 0.068 | -0.049 |

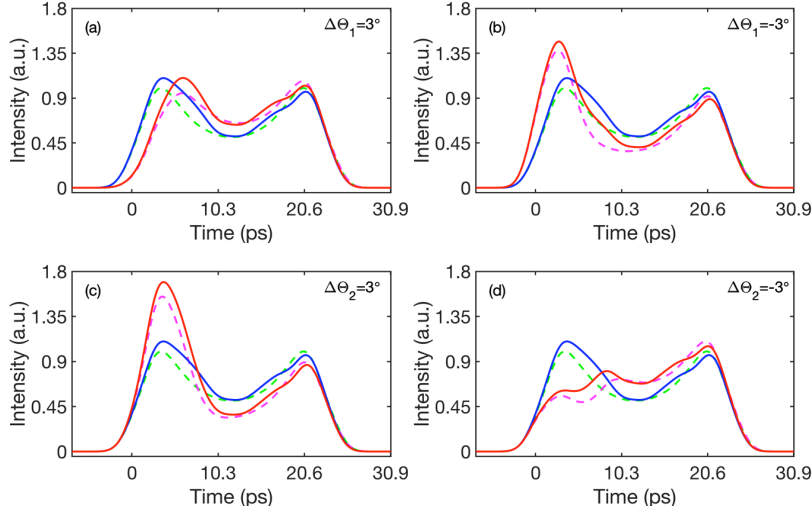

Fig. 1. Variations of shaped pulse temporal intensity profiles when changing the rotation angles of the 1st and 2nd BRs for ideal and non-ideal folded-type shapers with all the nominal BR phase delays set to $\pi$. $\Delta\Theta_n = \Theta_{n,tuned} - \Theta_{n,initial}$, where $n = 1, 2, \cdots, N$, and $\Theta_{n,initial}$ is the initial rotation angle of the $n-th$ BR defined in Eq. (1) for folded-type shaper and in Eq. (2) for fan-type shaper. The dashed lines are the shaped pulse temporal intensity profiles from ideal shapers, i.e., the time delays and phase delays of all BRs are the same with each other. The solid lines are the shaped pulse temporal intensity profiles from non-ideal shapers, i.e., the time delays and phase delays of the BRs are introduced with deviations as listed in Table 1. The green dashed lines and blue solid lines are the pulses from the folded-type shaper with initial elements rotation angles defined in Eq. (1) with $B_1 = 1$, and $B_2 = 90°$. The magenta dashed lines and red solid lines are the pulses from the shaper with only $\Theta_{1,or\ 2}$ changed by $3°$ or $-3°$, and all other parameter are the same with those for green dashed lines and blue solid lines, respectively. The input laser is Gaussian pulse with duration of 3ps (FWHM). The shapers have eight BRs with relative time delay ratio $\zeta$ of 1.

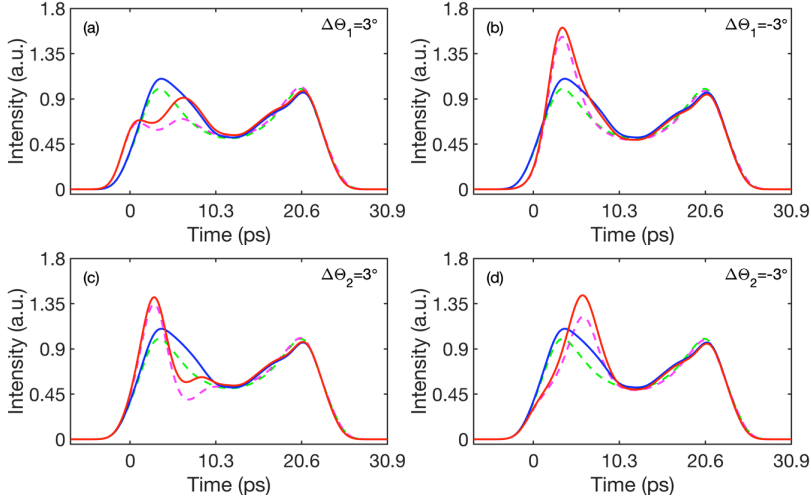

Fig. 2. Variation of shaped pulse temporal intensity profiles when changing the rotation angles of the 1st and 2nd BRs for ideal and non-ideal fan-type shapers with all the nominal BR phase delays set to 0 rad. The green dashed lines and blue solid lines are the pulses from the fan-type shaper with initial elements rotation angles defined in Eq. (2) with $B_1 = 1$, and $B_2 = 90°$. All other input parameters and meanings of the lines are the same with those in Fig. 1.

As shown in Figs. 1 and 2, though the shaped pulse profiles have significant changes when introducing the BRs time delay and phase delays with deviations listed in Table 1, the overall pulse profiles are still very similar to the corresponding pulse profiles from ideal shapers, e.g., green dashed lines vs. blue solid lines, and magenta dashed lines vs. red solid lines. The change trends of the shaped pulse profile when adjusting the BR rotation angles are the same between the ideal and non-ideal shapers (including both folded-type and fan-type shapers), e.g., green dashed lines vs. magenta dashed lines, and blue solid lines vs. red solid lines. This shows that the rules (in Table 1 and Table 2) of tuning BR rotation angles in ideal shapers still hold for non-ideal shapers introduced above, though the condition for the formation of $N + 1$ identical replicas may not be strictly satisfied for non-ideal shapers. In this paper, the configuration of $B_1 = 1, B_2 = 90^o$ is selected both for folded-type and fan-type shapers as an example. The characteristics of the folded-type or fan-type shaper are very similar for those four different configurations defined Eq. (1) and Eq. (2). The shaping algorithm, which would be presented below, is applicable to all these four different shaper configurations.

## 3. Adaptive shaping of folded-type variable shaper

The shaped pulse profile is determined by the rotation angles and phase delays of the BRs. No matter what the target profiles are, the shaped pulses obtain both high transmittance and high stability simultaneously at specific BRs phase delays ($\pi$ rad for folded-type shaper, 0 rad for fan-type shaper). A-cut birefringent crystals (e.g. YVO$_4$, BBO) are very suitable to be used as BRs. For laser with specific wavelength, the BR phase delay vs. temperature can be calibrated in advance. Therefore, the BR phase delays can be preset manually. The key to realize adaptive shaping is the adaptive optimization of the rotation angles of shaper's elements.

Besides the BRs rotation angles, the output polarizer's rotation angle $\Theta_p$ is also add to the optimization variables which are $\Theta = \{\Theta_1, \Theta_2, \cdots, \Theta_N, \Theta_p\}$. Figure 3 shows the changes of replicas and shaped pulse when adjusting the rotation angle of the output polarizer $\Theta_p$ in folded-type shaper with $B_1 = 1$, and $B_2 = 90°$. The shaper in Figs. 3 (a) and 3(b) contains even number ($N = 8$) of BRs. While for the shaper in Figs. 3(c) and 3(d), it contains odd number ($N = 9$) of BRs. As shown in Fig. 3, adjusting the output polarizer's rotation angle $\Theta_p$ can change the relative intensity between the $(N + 1) - th$ replica and all other replicas. Take the Fig. 3(a) for instance, when $\Theta_p$ is increased by $3°$ (i.e. $\Delta\Theta_p = \Theta_{p,tuned} - \Theta_{p,initial} = 3°$, where $\Theta_{p,initial}$ is the initial rotation angle of output polarizer defined in Eq. (1) for folded-type shaper and in Eq. (2) for fan-type shaper.), the 9th replica amplitude would decrease, while all other $N$ replicas' amplitudes would increase. The temporal profile of shaped pulse also changes correspondingly. When $\Theta_p$ is decreased by $3°$ (i.e. $\Delta\Theta_p = -3°$), as shown in Fig. 3(b), the 9th replicas and shape pulse change toward the opposite of that in Fig. 3(a). It shows that the effect of rotating output polarizer functions like "the $(N + 1) - th$ BR". Besides, the parity characteristic for rotating BRs in folded-type shaper is also hold for the output polarizer. As shown in Figs. 3(c) and 3(d), where there are odd number of BRs for the folded-type shaper, the effects of changing $\Theta_p$ on the replicas and shaped pulse profiles are the contrary of those in Figs. 3(a) and 3(b), respectively.

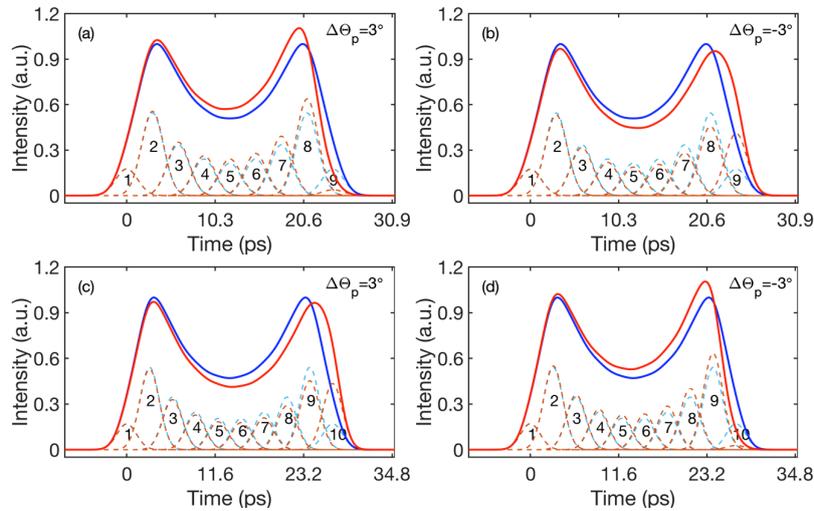

Fig. 3. Variations of the replicas and shaped pulses when changing the rotation angle of the output polarizer $\Theta_p$ in the folded-type shaper with $B_1 = 1$ and $B_2 = 90°$. The blue solid lines are the temporal intensity profiles of shaped pulses from the shaper with initial elements rotation angles as defined in Eq. (1). Its corresponding $N + 1$ replicas are represented with cyan dashed lines. The red solid lines are the temporal intensity profiles of the shaped pulses from the shaper with only $\Theta_p$ changed by $3°$ or $-3°$, and all other input parameters are the same with those for the pulse in blue solid line. Its corresponding $N + 1$ replicas are represented with brown dashed lines. The total number $N$ of BRs for figures (a) and (b) is eight, while for figures (c) and (d) is nine. The input laser is Gaussian pulse with 3ps (FWHM) duration. The relative time delay ratio $\zeta$ is 1. The BRs phase delays are set to $\pi$ rad.

In the practice, the temporal intensity distribution (or profile) of the output shaped pulse from the shaper can be measured by devices like streak camera, cross-correlator, etc. To realize adaptive optimization of the shaped pulse profile, $N + 1$ reference points of the complex temporal intensity profile of shaped pulse are selected to establish the connection with the elements of the shaper. The time coordinate of the $j - th$ reference point is the time coordinate corresponding to the peak intensity region of the $j - th$ replica pulse, where $j = 1,2,\cdots, N + 1$. Increasing the quantity of BRs would bring more replicas, therefore, leading higher resolution for the target pulse profile. For shaped pulse with complicated target profile, it would require large number of BRs to realize it.

Figures 4 and 5 are diagrams of the adaptive shaping algorithm for folded-type shaper. For the folded-type shaper described by Eq. (1), here $B_1 = 1, B_2 = 90^o$ are chosen as an example for demonstrating the shaping capability of the algorithm. For the remaining three configurations in Eq. (1), the algorithm works as well. The algorithm uses 6 parameters $S = \{\delta, \sigma, \beta, \xi, \eta, \rho\}$ to control the automatic shaping process. According to the $N + 1$ reference points of the current shaped pulse and target pulse, the algorithm automatically adjusts the rotation angles of the BRs and output polarizer of the shaper to tailor the output pulse approach to the target pulse. The target pulse's $N + 1$ reference points are denoted as $Q = \{Q_1, Q_2, \cdots, Q_{N+1}\}$. While for the shaped pulse outputted from the shaper, the $N + 1$ reference points are denoted as $R = \{R_1, R_2, \cdots, R_{N+1}\}$. Before comparing the shaped pulse with target pulse, the reference points are firstly normalized to the intensity of the 1st reference point, i.e., $R_1 = 1, Q_1 = 1$. To quantitatively evaluate the deviation of the shaped pulse from the desired pulse profiles, the shaping accuracy $\eta_{out}$ is introduced as cost function, which is the root-mean-square (RMS) error between $N + 1$ reference points of the shaped pulse and target pulse, as shown Eq. (4). The shaping algorithm would minimize the cost function to perform pulse shaping.

$$\eta_{out} = \sqrt{\frac{1}{N+1} \sum_{k=1}^{N+1} \left(\frac{\widetilde{R_k} - \widetilde{Q_k}}{\widetilde{Q_k}}\right)^2} \qquad (4)$$

Where $\widetilde{R_k} = R_k / \sum_{k=1}^{N+1} R_k$, $\widetilde{Q_k} = Q_k / \sum_{k=1}^{N+1} Q_k$.

$\delta$ define the adjustment range of the shaper's elements rotation angles. $\sigma$ is used to control the decreasing speed of $\delta$. As the shaping algorithm adjusts the shaper to tailor the shaped pulse profile to approach the target profile to a certain degree, the shaping error would not be further decreased efficiently unless using smaller adjustment range $\delta$, which makes the algorithm adjust the shaper in a finer way. $\delta$ has its maximum at the beginning of the iteration, which can be reduced in a way like Eq. (5).

$$\delta_{q+1} = \delta_q / \sigma \qquad (5)$$

Where $\delta_q$ is the current adjustment range of elements rotation angles, while $\delta_{q+1}$ is the updated new adjustment range of elements rotation angles.

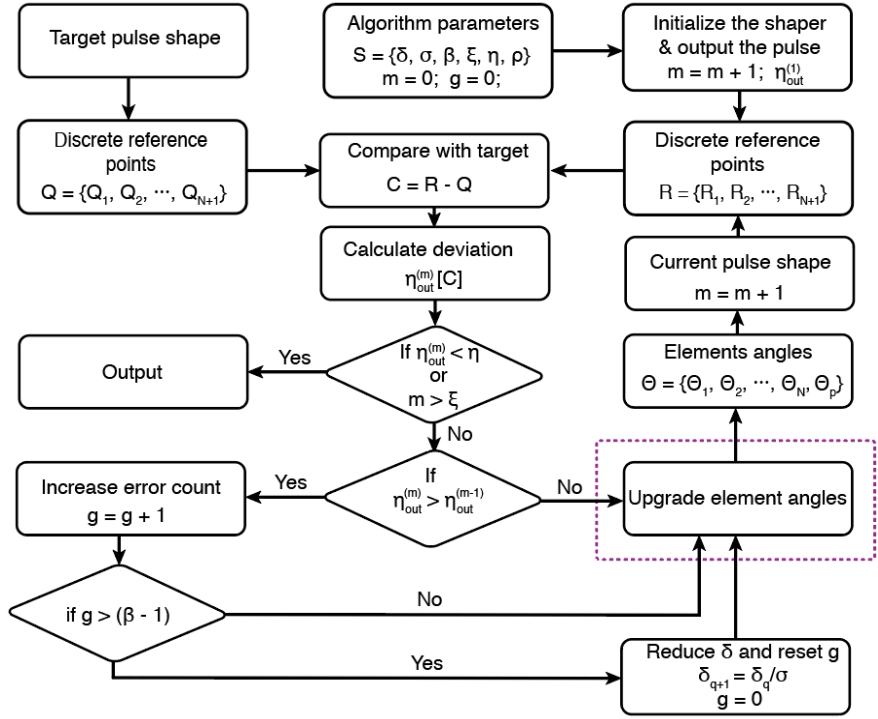

Fig. 4. Diagram of the shaping algorithm for folded-type shaper and fan-type shaper.

$\beta$ is the mix parameter. When adjusting the rotation angles of the shaper elements with adjustment range of $\delta_q$, if the iterations are not enough, the shaping error $\eta_{out}$ would be relatively large. At this point, if using Eq. (5) to update the $\delta$ to $\delta_{q+1}$, it would require more iterations to reduce the shaping error $\eta_{out}$, leading the algorithm's convergence speed decreases or even the algorithm begin to converge before the shaping error $\eta_{out}$ is larger than the target error $\eta$. Therefore, it needs mix parameter $\beta$ to make the algorithm adjust the shaper with $\delta_q$ adequately until the shaping error $\eta_{out}$ can't be reduced efficiently any more, then using Eq. (5) to update the $\delta$ to $\delta_{q+1}$. For the algorithm, once there are $\beta$ times that the shaping error $\eta_{out}^{(m)}$ bigger than the $\eta_{out}^{(m-1)}$, the algorithm would update the $\delta$ to $\delta_{q+1}$ with Eq. (5). $\xi$ is the parameter used to control the total number of iterations of the shaping algorithm. $\eta$ is the target error for the shaping algorithm. When the iteration count number $m$ is bigger than $\xi$ or shaping error $\eta_{out}$ less than the target error $\eta$, the algorithm would automatically stop and output the results. The variables that needs to be optimized for the shaper are the rotation angles of the $N$ BRs and the output polarizer, i.e., $\Delta\Theta = \{\Delta\Theta_1, \Delta\Theta_2, \cdots, \Delta\Theta_N, \Delta\Theta_p\}$. According to the calculation, the amplitude of the 1st replica pulse is mainly determined by the rotation angle of the 1st BR, as shown in Fig. A1 in the Appendix, while other elements have very small influence to it. The optimization of the 1st BR's rotation angle $\Delta\Theta_1$ is handled separately by presetting it to $\rho$ (i.e. $\Delta\Theta_1 = \rho$). The value of $\rho$ is set according to the shape of the desired pulse profile. For example, if the 1st replica's amplitude is much smaller than other replicas' amplitudes, $\rho$ should be set to bigger than 0 in order to reduce the 1st replica' amplitude, e.g. right sawtooth profile with peak intensity closer to the location of the $(N + 1) - th$ replicas. If 1st replica's amplitude is relatively bigger than other replicas' amplitudes, then $\rho$ should be set to less than 0 in order to increase the 1st replica' amplitude, e.g. left sawtooth profile with peak intensity closer to the location of the 1st replicas. To make the algorithm can both converge and generate shaped pulse with high efficiency, the value of $\rho$ should be appropriate chosen. One can first choose a small $|\rho|$ to ensure the algorithm can converge with shaping error $\eta_{out}$ less than target error $\eta$, then increasing $|\rho|$ to make the shaper's efficiency as large as possible. The comparison vector between the reference points of shaped pulse and target pulse is defined as:

$$C = R - Q \qquad (6)$$

If $B_2 = 0°$, the vector $C$ in Eq. (6) should be flipped to reverse the elements sequence, while for $B_2 = 90°$, the vector $C$ should be keep unchanged as in Eq. (5). This holds for both folded-type and fan-type shapers. When the shaping algorithm arrives to the dashed purple box in Fig. 4, the rotation angles of the shaper elements would be updated. The basic algorithm of the dashed purple box in Fig. 4 for folded-type shaper is presented in Fig. 5, which describes the general rules for adjusting the elements rotation angles in the shaping algorithm. Based on the tuning rules of the shaper's elements rotation angles described in Table 1, the rotation angle of the $n-th$ BR is updated according to Eq. (7) as follows:

$$\Theta_n^{(m+1)} = \Theta_n^{(m)} + D_n \cdot \delta \cdot C(n)/|C(n)| \qquad (7)$$

Where $n = 2, 3, \cdots, N$, and $D_n = (-1)^{n+1} B_1$, $m$ is the iteration count number. Since adjusting the $n-th$ BR's rotation angle $\Theta_n$ would simultaneously increase or decrease the $n-th$ and the $(n+1)-th$ replicas amplitudes, the $(n+1)-th$ BR's rotation angle $\Theta_{n+1}$ can keep unchanged in current iteration and the algorithm skips to adjust the $(n+2)-th$ BR's rotation angle $\Theta_{n+2}$ if $C(n) \cdot C(n+1) > 0$. However, if $C(n) \cdot C(n+1)$ is not bigger than 0, correction of $\Theta_{n+1}$ would lead the relative amplitude of the $(n+1)-th$ replica pulse deviate farther away from that in target output pulse, therefore the algorithm needs to continue to adjust the $(n+1)-th$ BR's rotation angle $\Theta_{n+1}$. If $C(n) = 0$, which is rare to occur, the $n-th$ BR's rotation angle $\Theta_n$ may also remain unchanged and the shaping algorithm may skip to adjust the $\Theta_{n+1}$ in current iteration. After all BRs' rotation angles are updated, the output polarizer's rotation angle $\Theta_p$ is adjusted according to Eq. (8).

$$\Theta_p^{(m+1)} = \Theta_p^{(m)} + D_{N+1} \cdot \delta \cdot C(N+1)/|C(N+1)| \qquad (8)$$

The updated rotation angles set $\Theta = \{\Theta_1, \Theta_2, \cdots, \Theta_N, \Theta_p\}$ outputted from algorithm in Fig. 5 is then used by the algorithm in Fig. 4 to adjust the shaper. For the first use of this shaping algorithm, the parameters $\delta, \beta$ should have relatively large value and parameter $\sigma$ should have relatively small value but bigger than 1 to ensure the algorithm can converge with shaping error $\eta_{out}$ less than target error $\eta$. Then it can gradually reduce $\beta$ and increase $\sigma$ to make the convergence speed of the algorithm as fast as possible. According to experience, with $\delta = 1°$, $\sigma \in (1,3)$, and $\beta \in (2\sim5)$, the shaping algorithm can converge with very fast speed and generate high fidelity shaped pulse in many cases. For shapers with different parameters, the algorithm control parameters $\delta, \beta$ and $\sigma$ can be chosen according to the method described above. For instance, to produce flattop pulse with folded-type shaper ($B_1 = 1, B_2 = 90°$) containing eight BRs, choosing $\zeta$ to be 0.8 for Gaussian input pulse with 2ps duration (FWHM). The BRs phase delays are set to $\pi$. Assuming the input parameters for the shaping algorithm are $\{\delta, \sigma, \beta, \rho\} = \{1°, 1.7, 3, -1°\}$ and the reference points set for the target pulse is $Q = \{0.67, 1, 1, 1, 1, 1, 1, 1, 0.67\}$. The algorithm has very fast convergence speed with shaping error $\eta_{out}$ drops to ~8% after 10 iterations, ~1% after 30 iterations and ~0.1% after 60 iterations.

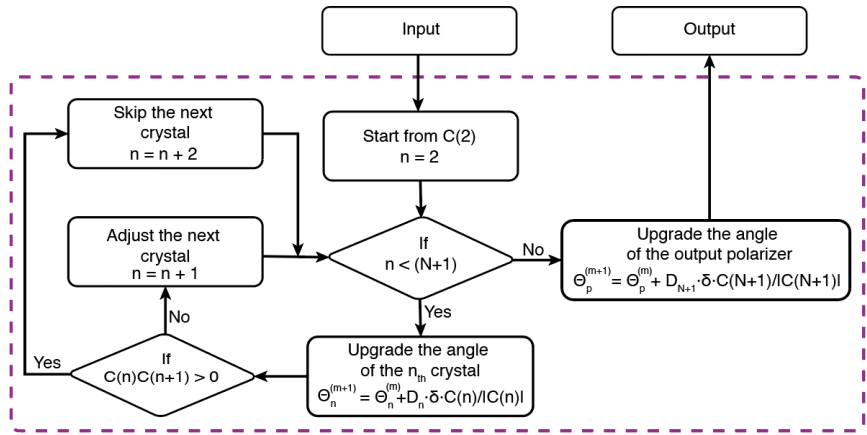

Fig. 5. Diagram of rotation angles update algorithm for folded-type shaper.

*3.1 Pulse shaping in ideal folded-type shaper*

In this section, the shaping algorithm described above is applied to a folded-type shaper containing 20 BRs to demonstrate the arbitrary temporal shaping with large number of BRs. The configuration of the folded-type shaper is $B_1 = 1, B_2 = 90^o$. There are three kinds of target pulses including smooth type, rippled type and type of micro-pulse train. The envelop of those target pulses is a very complicated shape, which is the combination of flattop, parabolic and triangular shape, and is named with "FPT" shape. Each target pulse is discretized into 21 reference points, whose relative intensities are listed in Table 4. The points 1~7 represent the flattop shape interval. For target pulse with smooth, the 1st reference point's value ($Q_1 = 0.67$) is smaller than that of 2nd, 3rd, ⋯ reference points' values. For smooth target pulse, the relative time delay ratio $\zeta$ is small (e.g. $\zeta < 1$ for Gaussian input pulse), then the optical interference between the adjacent replicas would be very strong. If $Q_1 = Q_2$, it would result in the shaped pulse intensity between the 1st and 2nd replicas larger than the intensities corresponding to the peak regions of the 1st and 2nd replicas, and the shaped pulse profile wouldn't be flat. Therefore, the $Q_1$ is set to less than $Q_2$ to eliminate such influence by decreasing the interference strength between 1st and 2nd replicas. However, for target pulse of rippled profile and micro-pulse train, the optical interference is much smaller than that in target pulse of smooth profile, therefore, the intensity of the 1st reference point $Q_1$ is set to be equal with $Q_2$ for flattop target pulse shape. The reference points 8~14 have their intensities satisfy the parabolic equation to represent the parabolic pulse shape. The reference points 15~21 represent the triangular pulse shape.

Table 4. Reference points intensities of FPT target pulse shape: (a) smooth type, (b) rippled type, (c) micro-pulse train

| Reference points | | Smooth | | Rippled | | Micro-pulse train | |
|---|---|---|---|---|---|---|---|
| 1 | 12 | 0.67 | 1.90 | 1.00 | 1.90 | 1.00 | 1.90 |
| 2 | 13 | 1.00 | 1.60 | 1.00 | 1.60 | 1.00 | 1.60 |
| 3 | 14 | 1.00 | 1.10 | 1.00 | 1.10 | 1.00 | 1.10 |
| 4 | 15 | 1.00 | 0.95 | 1.00 | 0.95 | 1.00 | 0.95 |
| 5 | 16 | 1.00 | 0.85 | 1.00 | 0.85 | 1.00 | 0.85 |
| 6 | 17 | 1.00 | 0.70 | 1.00 | 0.70 | 1.00 | 0.70 |
| 7 | 18 | 1.00 | 0.55 | 1.00 | 0.55 | 1.00 | 0.55 |
| 8 | 19 | 1.10 | 0.40 | 1.10 | 0.40 | 1.10 | 0.40 |
| 9 | 20 | 1.60 | 0.25 | 1.60 | 0.25 | 1.60 | 0.25 |
| 10 | 21 | 1.90 | 0.10 | 1.90 | 0.10 | 1.90 | 0.10 |
| 11 | — | 2.00 | — | 2.00 | — | 2.00 | — |

For the target pulse with complex profile described in Table 4, it is very hard to manually adjust the shaper for generating high fidelity shaped pulse because of the large number of optimization variables. In such situation, the shaping algorithm can be used to realize high fidelity pulse shaping. The input laser pulse has Gaussian profile with durations for the target pulses of smooth profile, rippled profile and type of micro-pulse train being 2ps, 1ps and 1ps, respectively. Meanwhile, the BR's relative time delay ratio $\zeta$ for target pulse with smooth profile, rippled profile and type of micro-pulse train are 0.8, 1.5 and 3.5, respectively. All the BRs phase delays are set to $\pi$ for the ideal folded-type to obtain high transmittance. The input parameters $\{\delta, \sigma, \beta, \eta\}$ for the shaping algorithm are set to $\{1°, 1.3, 5, 0.2\%\}$. The shaping error $\eta_{out}$ is required to be less than the target error $\eta$, therefore the iterations parameter $\xi$ can be set with a very big value. The algorithm would automatically stop and output the results once the shaping error $\eta_{out}$ is less than the target error $\eta$. With several tests, the 1st BR's rotation angle variation $\rho$ for target pulses with smooth profile, rippled profile and type of micro-pulse train are set to be $-0.86°, -0.84°$ and $-0.71°$, respectively.

With about 180 ~ 240 iterations, the shaping algorithm outputs the shaped pulses and the corresponding rotation angles adjustments of the shaper, as shown in Fig. 6 and Table 5. For all three different types of target pulses (including smooth profile, rippled profile and micro-pulse train), the shaping algorithm automatically stops and outputs the results when the shaping errors $\eta_{out}$ less than the target error $\eta = 0.2\%$. As shown in Fig. 6, no matter which type that target pulse profile is, the shaping algorithm can control the shaper to generate high fidelity shaped pulse with desired profile. Due to the discrepancies of the interference between different types of pulse profiles, the corresponding rotation angle adjustments required for the target pulses are different, as shown in Table 5. However, the shaping algorithm in this paper can adapt to such discrepancies and automatically optimize the element rotation angles of the shaper to produce the shaped pulse with high fidelity. According to the calculation, the transmittance of the folded-type and fan-type shapers is inversely related to the relative time delay ratios $\zeta$ and total number $N$ of BRs. Assuming an input unchirped Gaussian pulse with duration of 1ps (FWHM), and the shaper contains seven BRs with $\zeta = 0.4$. The laser transmittance can be above 50% for folded-type shaper with $\pi$ rad BRs phase delays and for fan-type shaper with 0 rad BRs phase delays.

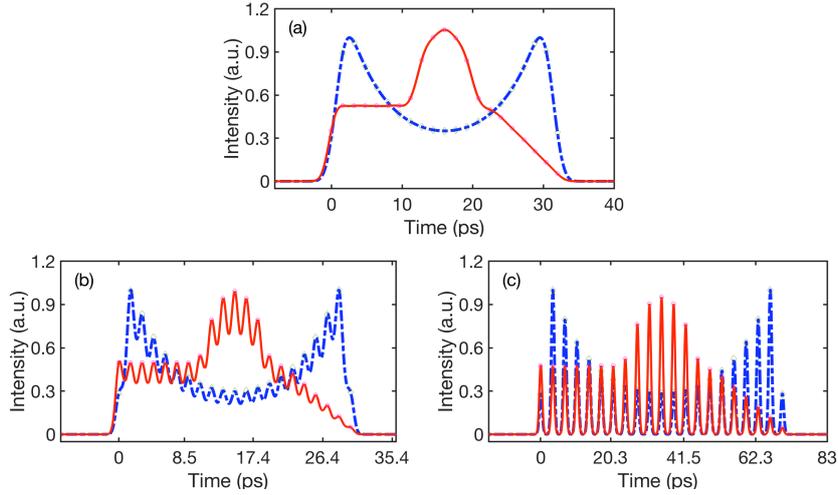

Fig. 6. Three different simulated shaped pulses including (a) smooth profile, (b) rippled profile, (c) micro-pulse train, which are generated by the shaping algorithm in ideal folded-type shaper ($B_1 = 1$, and $B_2 = 90°$) for picosecond laser pulse. The shaping efficiency for the three different pulses are about (1) 12.1%, (b) 7.2%, and (c) 5.1%. The blue dash-dot lines are the pulses from the shaper with initial elements rotation angles defined in Eq. (1). Its corresponding 21 reference points are shown with green diamond markers. The red solid lines are the shaped pulses generated by the shaping algorithm. Its corresponding 21 reference points are shown with magenta circle markers. The input parameters $\{\delta, \sigma, \beta, \eta\}$ for the shaping algorithm are set to $\{1°, 1.3, 5, 0.2\%\}$. The input laser is unchirped Gaussian pulse with duration (FWHM) of (a) 2ps, (b) 1ps, (c) 1ps. The shaper contains 20 BRs with relative time delay ratios of the BR are (a) $\zeta = 0.8$, (b) $\zeta = 1.5$, (c) $\zeta = 3.5$. The phase delays of the BRs are $\pi$ rad for high transmittance. For the given BRs phase delays, the shaped pulse profiles for input pulses with different wavelengths (e.g. 1064nm, 532nm, 266nm, ⋯) are the same with each other when neglecting the material dispersion.

Table 5. Elements rotation angle variations $\Delta\Theta_n$ (Deg) for different types of shaped pulses showed in Fig. 6 (red solid lines) including (a) smooth profile, (b) rippled profile, (c) micro-pulse train, where $N = 20$, $n = 1, 2, \cdots, N, N+1$ and $\Delta\Theta_{N+1} = \Delta\Theta_p$.

| Element number ($n$) | | Smooth | | Rippled | | Micro-pulse train | |
|---|---|---|---|---|---|---|---|
| 1 | 12 | -0.860 | -2.683 | -0.840 | -0.855 | -0.710 | -0.697 |
| 2 | 13 | -2.461 | -3.013 | -2.024 | -2.113 | -1.892 | -2.005 |
| 3 | 14 | -1.189 | -3.931 | -1.214 | -1.791 | -1.091 | -1.600 |
| 4 | 15 | -2.300 | -3.023 | -1.656 | -1.619 | -1.519 | -1.464 |
| 5 | 16 | -1.402 | -4.320 | -1.552 | -2.265 | -1.453 | -2.080 |
| 6 | 17 | -2.166 | -2.602 | -1.336 | -1.062 | -1.204 | -0.899 |
| 7 | 18 | -1.638 | -4.712 | -1.822 | -2.828 | -1.732 | -2.643 |
| 8 | 19 | -2.314 | -2.141 | -0.994 | -0.506 | -0.831 | -0.324 |
| 9 | 20 | -2.928 | -5.179 | -2.618 | -3.415 | -2.512 | -3.226 |
| 10 | 21 | -2.331 | -3.903 | -0.597 | -2.109 | -0.438 | -1.909 |
| 11 | — | -3.190 | — | -2.676 | — | -2.575 | — |

*3.2 Pulse shaping in non-ideal folded-type shaper*

Based on the section 3.1, in this section, the shaping algorithm is further applied to non-ideal folded-type shaper, which has nonidentical BRs times delays and phase delays, for arbitrary optical waveform generation. The nominal parameters of the non-ideal folded-type shaper are the same with those in ideal folded-type shaper introduced in section 3.1. For non-ideal folded-type shaper, its BRs time delays and phase delays are introduced with random deviations as shown in Eq. (9) and Eq. (10).

$$\Delta\tau = -W_1 + 2W_1 \cdot \boldsymbol{M}_1 \qquad (9)$$

$$\Delta\varphi = -W_2 + 2W_2 \cdot \boldsymbol{M}_2 \qquad (10)$$

Where $W_1$ and $W_2$ are the tolerances for BRs time delays and phase delays, respectively. $M_1$ and $M_1$ are the random number vectors between 0 and 1, which are listed in Table 6 below.

Table 6. Random number vectors between 0 and 1 for $W_1$ and $W_2$

| BR number ($n$) | 1 | 2 | 3 | 4 | 5 | 6 | 7 | 8 | 9 | 10 |
| | 11 | 12 | 13 | 14 | 15 | 16 | 17 | 18 | 19 | 20 |
|---|---|---|---|---|---|---|---|---|---|---|
| $M_1$ | 0.162 | 0.794 | 0.311 | 0.529 | 0.166 | 0.602 | 0.263 | 0.654 | 0.689 | 0.748 |
| | 0.451 | 0.084 | 0.229 | 0.913 | 0.152 | 0.826 | 0.538 | 0.996 | 0.078 | 0.443 |
| $M_2$ | 0.656 | 0.036 | 0.849 | 0.934 | 0.679 | 0.758 | 0.743 | 0.392 | 0.655 | 0.171 |
| | 0.706 | 0.032 | 0.277 | 0.046 | 0.097 | 0.823 | 0.695 | 0.317 | 0.950 | 0.034 |

With about ~200 iterations, the shaping algorithm automatically stops and outputs the results when shaping error $\eta_{out}$ less than the target error $\eta = 0.2\%$ for three different types of target pulses. The relative intensities of 21 reference points for target pulses are presented in Table 4. The shaped pulses and the corresponding elements rotation angles adjustments are presented in Fig. 7 and Table 7, respectively. As shown in Fig. 7, the deviations introduced to the BRs time delays and phase delays can distort the shaped pulse profile. However, the shaping algorithm for folded-type shaper can adapt to such deviations in the BRs by adjusting the rotation angles of the shaper's elements to make the shaped pulse approach to the desired profile. With the time delay deviations introduced to the BRs, the time coordinates of the sub-wave crests of rippled pulse would deviate from those in ideal shapers. Similar effects would also happen for target pulse of micro-pulse train type, for which the time coordinates of the micro-pulses' peaks would deviate from those in ideal shapers due to time delay deviations introduced to the BRs. One can move the time coordinates of the reference points to make them matched with the new sub-wave crests of rippled output pulse or new micro-pulses' peaks of the pulse train, as shown in Figs. 7(b) and 7(c). Though the time coordinates of the reference points deviate from those in ideal shapers described in section 3.1 above, the rotation angle tuning rules of the shaper still hold for such situation as long as the deviations aren't very big.

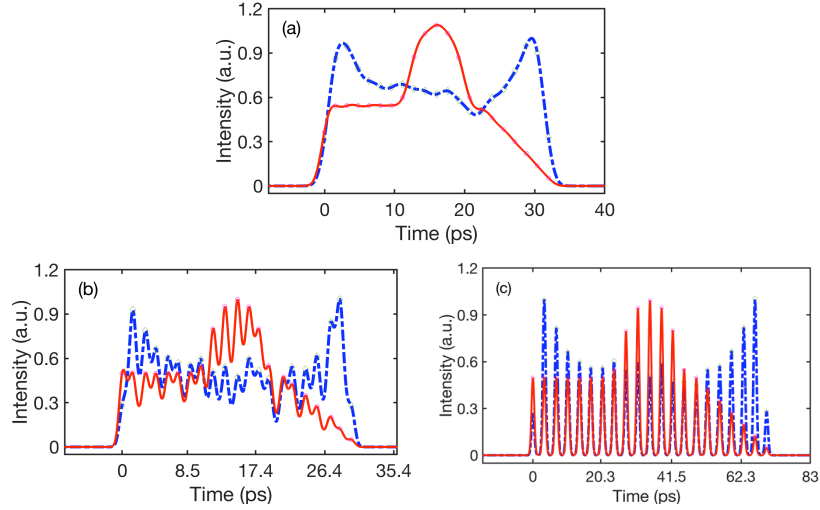

Fig. 7. Three different shaped pulses including (a) smooth profile, (b) rippled profile, (c) micro-pulse train, which are generated by the shaping algorithm in non-ideal folded-type shaper ($B_1 = 1$, and $B_2 = 90°$). The shaping efficiency for the three different pulses are about (1) 12.6%, (b) 7.7%, and (c) 5.4%. The BRs of the shaper are introduced with deviations as shown in Eq. (9), Eq. (10) and Table 6. The tolerances $\{W_1, W_2\}$ for target pulses of (a) smooth profile, (b) rippled profile, and (c) micro-pulse train are $\{0.05ps, 0.1\pi\}$, $\{0.15ps, 0.1\pi\}$, and $\{0.15ps, 0.1\pi\}$, respectively. The nominal relative time delay ratios of the BRs are (a) $\zeta = 0.8$, (b) $\zeta = 1.5$, (c) $\zeta = 3.5$. The nominal phase delays of the BRs are $\pi$ rad. All other input parameters and meanings of the lines and markers are the same with those in Fig. 6 for folded-type shaper.

Table 7. Elements rotation angle variations $\Delta\Theta_n$ (Deg) for different types of shaped pulses showed in Fig. 7 (red solid lines) including (a) smooth profile, (b) rippled profile, (c) micro-pulse train, where $N = 20$, $n = 1, 2, \cdots, N, N+1$ and $\Delta\Theta_{N+1} = \Delta\Theta_p$.

| Element number ($n$) | | Smooth | | Rippled | | Micro-pulse train | |
|---|---|---|---|---|---|---|---|
| 1 | 12 | -0.860 | -3.155 | -0.840 | -1.959 | -0.710 | -1.623 |
| 2 | 13 | -2.413 | -3.215 | -2.119 | -2.058 | -2.009 | -1.826 |
| 3 | 14 | -1.143 | -4.313 | -1.103 | -2.569 | -0.919 | -2.339 |
| 4 | 15 | -2.253 | -3.461 | -1.944 | -1.835 | -1.806 | -1.598 |
| 5 | 16 | -1.210 | -4.921 | -1.185 | -2.993 | -1.022 | -2.760 |
| 6 | 17 | -2.250 | -3.292 | -1.872 | -1.582 | -1.729 | -1.227 |
| 7 | 18 | -1.366 | -5.432 | -1.239 | -3.460 | -1.044 | -3.199 |
| 8 | 19 | -2.619 | -2.943 | -1.799 | -1.093 | -1.638 | -0.757 |
| 9 | 20 | -2.760 | -6.018 | -2.173 | -4.093 | -1.811 | -3.702 |
| 10 | 21 | -2.629 | -4.809 | -1.499 | -2.803 | -1.359 | -2.389 |
| 11 | — | -3.125 | — | -2.360 | — | -2.044 | — |

## 4. Adaptive shaping of fan-type variable shaper

Similar to the folded-type shaper, the fan-type shaper also has its high transmittance appear at a specific BR phase delay for different pulse profiles. For fan-type shaper, this special BR phase delay is 0 rad. The phase delays of the BRs in the fan-type shaper can be preset manually, which is similar to that in folded-type shaper described above. The key for adaptive shaping with fan-type shaper is the adaptive optimization of the shaper's elements rotation angles, which are $\Theta = \{\Theta_1, \Theta_2, \cdots, \Theta_N, \Theta_p\}$.

Figure 8 shows the variations of replicas and shaped pulses when changing the output polarizer's rotation angle $\Theta_p$ in fan-type shaper with $B_1 = 1$ and $B_2 = 90°$. The total number of BRs is even number ($N = 8$) for the shaper of Figs. 8 (a) and 8(b) and is odd number ($N = 9$) for the shaper of Figs. 8 (c) and 8(d). As shown in Fig. 8, the effect of changing output polarizer rotation angle $\Theta_p$ in the fan-type shaper also functions like "the $(N + 1) - th$ BR". However, for fan-type shaper, the variations of the replicas and shaped pulse when changing $\Theta_p$ is very different from what happens when changing the rotation angles of the BRs (e.g. Fig. A2 in the Appendix below). Taking the Fig. 8(a) for instance, when $\Theta_p$ is increased by $3°$(i.e. $\Delta\Theta_p = 3°$), the amplitude of the 9th replica would increase and all other replicas' amplitudes would decrease. The shaped pulse would also vary correspondingly. When $\Theta_p$ is decreased by $3°$ (i.e. $\Delta\Theta_p = -3°$), as shown in Fig. 8 (b), the replicas and shaped pulse change towards the opposite of what happens in Fig. 8(a). This are very similar to the cases in Figs. 3 (c) and 3(d) for folded-type shaper. However, there no parity characteristics for the output polarizer in fan-type shaper. As shown in Figs. 8(c) and 8(d), for which the fan-type shaper contains odd number ($N = 9$) of BRs, the change trends of the replicas and shaped pulses profiles when adjusting the output polarizer rotation angle $\Theta_p$ are the same with those in Figs. 8 (a) and 8(b).

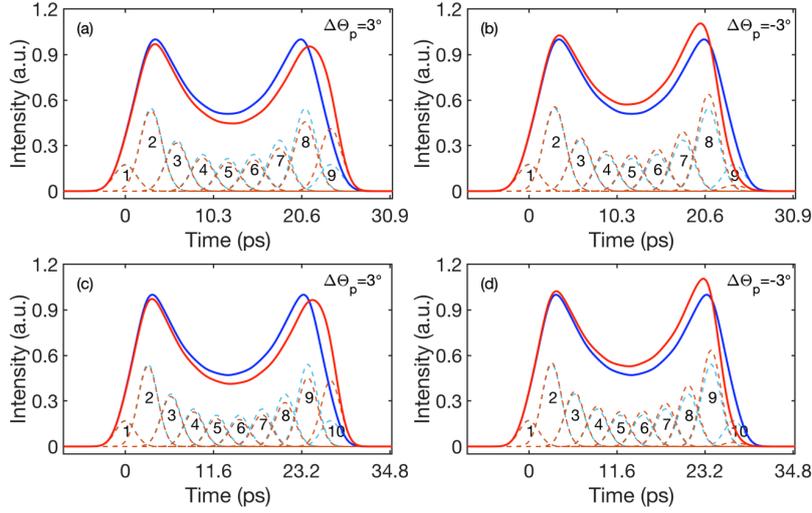

Fig. 8. Variations of the replicas and shaped pulses when changing the rotation angle of the output polarizer $\Theta_p$ in the fan-type shaper with $B_1 = 1$ and $B_2 = 90°$. The blue solid lines are the temporal intensity profiles of shaped pulses from the shaper with initial elements rotation angles as defined in Eq. (2). Its corresponding $N + 1$ replicas are represented with cyan dashed lines. The red solid lines are the temporal intensity profiles of the shaped pulses from the shaper with only $\Theta_p$ changed by $3°$ or $-3°$, and all other input parameters are the same with those for pulse in blue solid lines. Its corresponding $N + 1$ replicas are represented with brown dashed lines. The total number of BRs for figures (a) and (b) is eight, while for figures (c) and (d) is nine. The input laser is Gaussian pulse with 3ps (FWHM) duration. The relative time delay ratio $\zeta$ is 1. The BRs phase delays are set to 0 rad.

The shaping algorithm for fan-type shaper is diagramed in Fig. 4 and Fig. 9. The elements rotation angles update module——the dashed purple box in Fig. 4 for fan-type shaper is different from that for folded-type shaper. The basic algorithm of the dashed purple box in Fig. 4 for fan-type shaper is presented in Fig. 9, which describes the general rules for tuning the elements rotation angles in the shaping algorithm. Besides, only five parameters $S = \{\delta, \sigma, \beta, \xi, \eta\}$ are used by the shaping algorithm to control the shaping process for fan-type shaper. The meanings of these five parameters are the same with those for folded-type shaper introduced above. For fan-type shaper, the adjustment of the output polarizer rotation angle $\Theta_p$ can be omitted (i.e., $\Theta_p = 0°$), whereas the rotation angle of the 1st BR can be included into the iteration process of the algorithm. As shown in Eq. (11), the update of the BRs' rotation angles can be express as:

$$\Theta_{N+1-n}^{(m+1)} = \Theta_{N+1-n}^{(m)} + D \cdot \delta \cdot C(N + 2 - n)/|C(N + 2 - n)| \qquad (11)$$

Where $n = 1,2,\cdots,N$, and $D = B_1$. $B_1 = 1$, and $B_2 = 90°$ are chosen for fan-type shaper as an example to demonstrate the shaping capability of the algorithm, which works as well for the remaining three configurations in Eq. (2). Since adjusting the $(N + 1 - n) - th$ BR's rotation angle $\Theta_{N+1-n}$ would simultaneously change the amplitudes of the $(N + 2 − n) − th$ and $(N + 1 − n) − th$ replicas, the $(N − n) − th$ BR's rotation angle $\Theta_{N-n}$ can keep unchanged in current iteration if $C(N + 2 − n) \cdot C(N + 1 − n) < 0$, and at this point, the algorithm skips to adjust the $(N − 1 − n) − th$ BR's rotation angle $\Theta_{N-1-n}$. If $C(N + 2 − n) \cdot C(N + 1 − n)$ is not less than 0, the algorithm needs to continue to adjust the $(N − n) − th$ BR's rotation angle $\Theta_{N-n}$, since correction of $\Theta_{N+1-n}$ would lead the relative amplitude of the $(N + 1 − n) − th$ replica pulse deviate farther away from that in target output pulse. If $C(N + 2 − n) = 0$, which is rare to occur, the $(N + 1 − n) − th$ BR's rotation angle $\Theta_{N+1-n}$ may also remain unchanged and the shaping algorithm can skip to adjust the $\Theta_{N-n}$ in current iteration. Different from the shaping algorithm for folded-type shaper, which skips to adjust the $(n + 2) − th$ BR under $C(n) \cdot C(n + 1) > 0$, the shaping algorithm for fan-type shaper skips to adjust the $(N − 1 − n) − th$ BR under $C(N + 2 − n) \cdot C(N + 1 − n) < 0$. Such difference is caused by the different element rotation angle tuning rules between folded-type and fan-type shapers. After all the rotation angles of the BRs are updated, the algorithm in Fig. 9 output the new rotation angles to the algorithm in Fig. 4 to adjust the fan-type shaper. The principles of choosing values for the control parameters $\{\delta, \sigma, \beta\}$ of the shaping algorithm for fan-type shaper are the same with those in folded-type shaper described above.

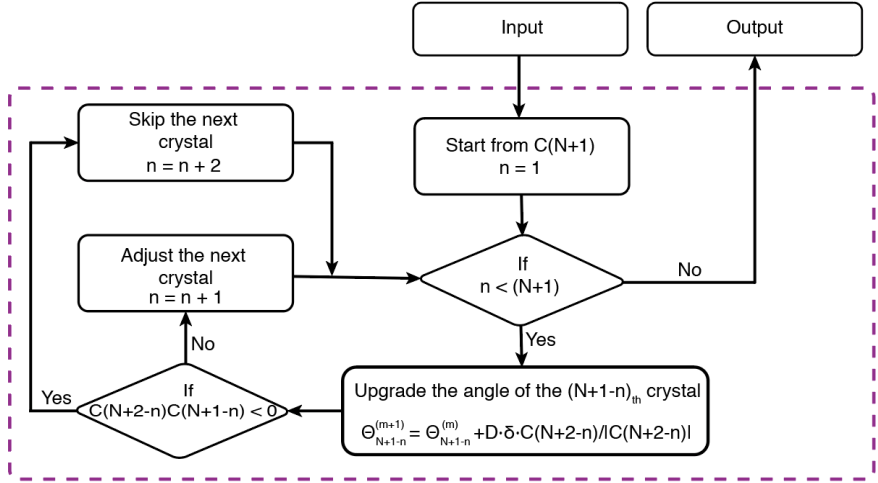

Fig. 9. Diagram of rotation angles update algorithm for fan-type shaper.

### 4.1 Pulse shaping in ideal fan-type shaper

In this section, the shaping algorithm is applied to a fan-type shaper containing 20 BRs. The fan-type shaper's configuration is selected to be $B_1 = 1, B_2 = 90^o$ as an example. The target pulses are the same with those in section 3.1, which includes smooth profile, rippled profile and type of micro-pulse train. The relative intensities of 21 reference points for target pulses are presented in Table 4. The Gaussian input pulses' durations for the target pulses with smooth profile, rippled profile and micro-pulse train are 2ps, 1ps and 1ps, respectively. Meanwhile, the relative time delay ratio $\zeta$ of the BR for target pulses with smooth profile, rippled profile and micro-pulse train are 0.8, 1.5 and 3.5, respectively. To obtain high transmittance, all the BRs phase delays are set to 0 rad for the fan-type shaper. The input parameters $\{\delta, \sigma, \beta, \eta\}$ for the shaping algorithm are $1°, 1.3, 5$ and $0.2\%$, respectively. With about 200 ~ 230 iterations, the shaping algorithm for fan-type shaper stops and outputs the results automatically with shaping error $\eta_{out}$ less than the target error $\eta = 0.2\%$ for all three types of target pulses. The shaped pulses and corresponding rotation angles adjustments outputted by the algorithm are shown in Fig. 10 and Table 8, respectively. Similar with the shaping algorithm for folded-type shaper in section 3.1, the shaping algorithm for the fan-type shaper can also adapt to the discrepancies of the three different types of target pulses by adjusting the rotation angles of the BRs in the shaper, and generate shaped pulses with high fidelity.

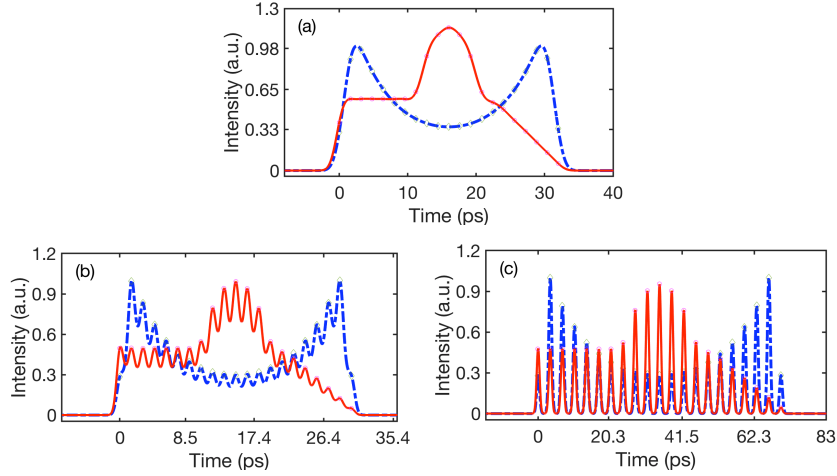

Fig. 10. Three different shaped pulses including (a) smooth profile, (b) rippled profile, (c) micro-pulse train, which are generated by the shaping algorithm in ideal fan-type shaper ($B_1 = 1$, and $B_2 = 90°$) for picosecond laser pulse. The shaping efficiency for the three different pulses are about (1) 13.2%, (b) 7.2%, and (c) 5.1%. The blue dash-dot lines are the pulses from the shaper with initial elements rotation angles defined in Eq. (2). Its corresponding 21 reference points are shown with green diamond markers. The red solid lines are the shaped pulses generated by the shaping algorithm. Its corresponding 21 reference points are shown with magenta circle markers. The input parameters $\{\delta, \sigma, \beta, \eta\}$ for the shaping algorithm are set to $\{1°, 1.3, 5, 0.2\%\}$. The input laser is unchirped Gaussian pulse with duration (FWHM) of (a) 2ps, (b) 1ps, (c) 1ps. The shaper contains 20 BRs with relative time delay ratios of the BR are (a) $\zeta = 0.8$, (b) $\zeta = 1.5$, (c) $\zeta = 3.5$. The phase delays of the BRs are 0 rad for high transmittance.

Table 8. Elements rotation angle variations $\Delta\Theta_n$ (Deg) for different types of shaped pulses showed in Fig. 10 (red solid lines) including (a) smooth profile, (b) rippled profile, (c) micro-pulse train, where $N = 20$, $n = 1, 2, \cdots, N$.

| BR number ($n$) | | Smooth | | Rippled | | Micro-pulse train | |
|---|---|---|---|---|---|---|---|
| 1 | 11 | 1.082 | 6.421 | 0.844 | 6.391 | 0.712 | 6.318 |
| 2 | 12 | -0.161 | 8.275 | -0.317 | 8.321 | -0.451 | 8.265 |
| 3 | 13 | -0.903 | 9.817 | -1.090 | 9.678 | -1.224 | 9.630 |
| 4 | 14 | -1.322 | 9.906 | -1.484 | 10.091 | -1.618 | 10.086 |
| 5 | 15 | -1.353 | 9.876 | -1.518 | 9.984 | -1.646 | 9.993 |
| 6 | 16 | -1.107 | 9.290 | -1.220 | 9.396 | -1.351 | 9.407 |
| 7 | 17 | -0.474 | 8.108 | -0.639 | 8.239 | -0.758 | 8.255 |
| 8 | 18 | 0.097 | 6.388 | 0.303 | 6.504 | 0.210 | 6.528 |
| 9 | 19 | 2.105 | 4.087 | 2.049 | 4.196 | 1.960 | 4.214 |
| 10 | 20 | 4.144 | 1.202 | 4.190 | 1.301 | 4.106 | 1.315 |

*4.2 Pulse shaping in non-ideal fan-type shaper*

In this section, the shaping algorithm is further applied to non-ideal fan-type shaper ($B_1 = 1, B_2 = 90°$). The deviations introduced to the time delays and phase delays of the BRs are shown in Eq. (9), Eq. (10) and Table 6. The tolerances $\{W_1, W_2\}$ for target pulses of smooth profile, rippled profile, and micro-pulse train are $\{0.05ps, 0.1\pi\}$, $\{0.15ps, 0.1\pi\}$, and $\{0.15ps, 0.1\pi\}$, respectively. The input laser pulses and the corresponding nominal relative time delay ratio $\zeta$ values for three different target pulses are the same with those in section 4.1 for ideal fan-type shaper. The nominal phase delays of BRs are 0 rad. With about ~200 iterations, the shaping algorithm stops and outputs the results (as shown in Fig. 11 and Table 9) automatically with $\eta_{out}$ less than the target error $\eta = 0.2\%$ for all three types of target pulses. The relative intensities of 21 reference points for target pulses are listed in Table 4. Similar to the folded-type shaper, the shaping algorithm is also applicable to non-ideal fan-type shaper to realize arbitrary temporal shaping. The time coordinates for the sub-wave crests of rippled pulse and for the micro-pulses' peaks of the pulse train would deviate from those in ideal shapers, because of the deviations introduced to BRs time delays. One can move the time coordinates of the reference points to the new coordinates corresponding the sub-wave crests of rippled output pulse or the micro-pulses' peaks of the pulse train.

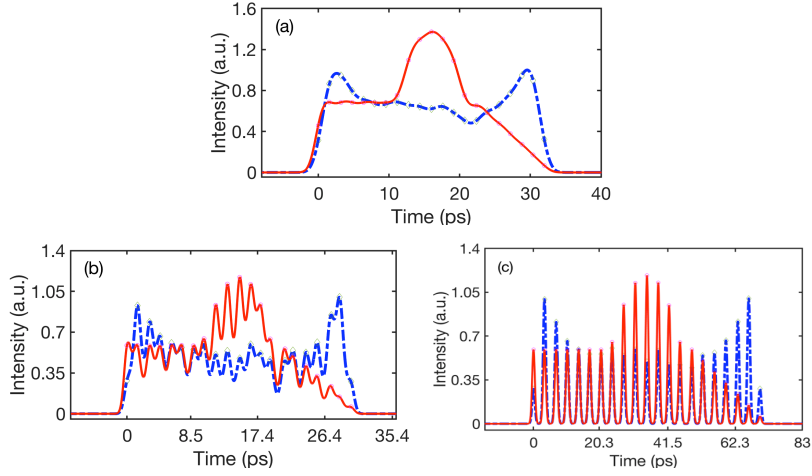

Fig. 11. Three different shaped pulses including (a) smooth profile, (b) rippled profile, (c) micro-pulse train, which are generated by the shaping algorithm in non-ideal fan-type shaper ($B_1 = 1$, and $B_2 = 90°$). The shaping efficiency for the three different pulses are about (1) 15.9%, (b) 8.9%, and (c) 6.4%. The BRs of the shaper are introduced with deviations as shown in Eq. (9), Eq. (10) and Table 6. The tolerances $\{W_1, W_2\}$ for target pulses of (a) smooth profile, (b) rippled profile, and (c) micro-pulse train are $\{0.05 ps, 0.1\pi\}$, $\{0.15 ps, 0.1\pi\}$, and $\{0.15 ps, 0.1\pi\}$, respectively. The nominal relative time delay ratios of the BRs are (a) $\zeta = 0.8$, (b) $\zeta = 1.5$, (c) $\zeta = 3.5$. The nominal phase delays of the BRs are 0 rad. All other input parameters and meanings of the lines and markers are the same with those in Fig. 10 for fan-type shaper.

Table 9. Elements rotation angle variations $\Delta\Theta_n$ (Deg) for different types of shaped pulses showed in Fig. 11 (red solid lines) including (a) smooth profile, (b) rippled profile, (c) micro-pulse train, where $N = 20$, $n = 1, 2, \cdots, N$.

| BR number ($n$) | | Smooth | | Rippled | | Micro-pulse train | |
|---|---|---|---|---|---|---|---|
| 1 | 11 | 1.321 | 5.338 | 1.157 | 5.720 | 1.051 | 5.330 |
| 2 | 12 | 0.416 | 6.653 | 0.338 | 7.088 | 0.239 | 6.819 |
| 3 | 13 | -0.032 | 8.143 | -0.072 | 8.204 | -0.204 | 8.151 |
| 4 | 14 | -0.091 | 8.311 | -0.165 | 8.603 | -0.284 | 8.632 |
| 5 | 15 | -0.020 | 8.714 | -0.112 | 8.743 | -0.180 | 8.831 |
| 6 | 16 | 0.103 | 8.199 | 0.072 | 8.261 | 0.051 | 8.408 |
| 7 | 17 | 0.403 | 7.395 | 0.303 | 7.461 | 0.301 | 7.508 |
| 8 | 18 | 0.214 | 5.881 | 0.580 | 6.046 | 0.602 | 6.025 |
| 9 | 19 | 1.752 | 3.842 | 1.986 | 3.995 | 1.876 | 3.928 |
| 10 | 20 | 3.264 | 1.053 | 3.693 | 1.187 | 3.416 | 1.205 |

## 5. Femtosecond laser pulse shaping with material dispersion

The shaping process above neglects the influence of material dispersion in the shaper. This is appropriate for laser pulse with narrow bandwidth such as picosecond lasers, since the material dispersion has very small impact on it. For laser pulses with wide bandwidth like femtosecond (fs) lasers, the effect of material dispersion may be significant in these cases. The relative time delay ratio $\zeta$ of BR is related with the input pulse duration. Therefore, laser pulse with shorter duration would require shorter BR length for shaping, and the material dispersion would also decrease correspondingly. This feature would help to mitigate the effect of material dispersion on the femtosecond laser pulses. The shaping algorithm in this paper is based on the effects of adjusting shaper's elements rotation angles on the shaped pulse profile. In this section, the shaping algorithm would be applied to shaping femtosecond laser pulse considering the material dispersion of the shaper. According to the calculation, it is found that the existence of material dispersion would reduce shaper's transmittance, especially for the output shaped pulses of smooth or rippled profiles. While for the transmittance of the output shaped pulse with type of micro-pulse train, the material dispersion has little impact on it. $\alpha - BBO$ crystals of a-cut are chosen to be BRs for shaping ultraviolet femtosecond laser pulse as an example. The input laser is unchirped Gaussian pulse with duration of 80fs (FWHM) and center wavelength of 266nm. The refractive index for o light and e light in the $\alpha - BBO$ crystal can be described by the Sellmeier equation as shown in Eq. (12) and Eq. (13).

$$n_o^2 = 2.7471 + \frac{0.01878}{\lambda^2 - 0.01822} - 0.01354\lambda^2 \qquad (12)$$

$$n_e^2 = 2.37153 + \frac{0.01224}{\lambda^2 - 0.01667} - 0.01516\lambda^2 \qquad (13)$$

Where $\lambda$ is light wavelength in $\mu m$. With the influence of material dispersion, the replica pulses would change as they propagate through the shaper. For the shaped pulse generated by the shaper, it can first calculate the complex amplitude of the light spectral component after the shaper, then using Fourier transform to obtain the temporal intensity distribution and phase of the output shaped pulse. There are three types of target pulses, which are smooth profile, rippled profile and micro-pulse train, with the corresponding relative time delay ratio $\zeta$ being 0.62, 1.1 and 2, respectively. Both the folded-type and fan-type shapers contain 20 BRs. The relative intensities of the 21 reference points for three different types of target pulses are listed in Table 4. However, the relative intensity of the 1st reference point for target pulse of smooth profile is 0.52. The input parameters for the algorithm are $\{\delta, \sigma, \beta, \eta\} = \{1°, 1.4, 5, 0.2\%\}$. In general, the durations of the wideband replica pulses and the output shaped pulse would be broadened by the material dispersion. Therefore, the time coordinates of the $N + 1$ reference points would be different from those in the ideal shaper with no material dispersion. For target pulses of rippled and micro-pulse train type, the structure of the $N + 1$ replicas are obvious and the time coordinates of the $N + 1$ reference points can be shifted to the new sub-wave crests of rippled pulse or the micro-pulses' peaks of the pulse train. However, for target pulse with smooth profile, the structure of the $N + 1$ replicas are invisible. To establish the connection between the output pulse and the elements of the shaper, it may first adjust the BRs phase delays to induce destructive interference. The Gaussian like laser pulse has its maximum value at the pulse's center, which bears the smallest influence of the destructive interference. Therefore, the structure of the $N + 1$ replicas would appear at this point. Figure 12 shows the calculated temporal profiles of the output pulse from the folded-type shaper with BRs phase delays of $0.1\pi$. One can see that there are $N + 1$ sub-wave crests or micro-pulses presented in the output pulse. The time coordinates of the $N + 1$ reference points may be shift to near the sub-wave crests or micro-pulses' peaks of the output pulses, as shown in Fig. 12. With this method, the time coordinates of the $N + 1$ reference points are updated. With about 150~190 iterations, the shaping algorithm automatically stops and outputs the results with $\eta_{out}$ less than the target value $\eta = 0.2\%$ for all three types of target pulses. The shaped pulses and the corresponding rotation angles adjustments are presented in Fig. 13, Table 10, Fig. 14 and Table 11. The output shaped pulse of smooth profile has the largest efficiency because of the small relative time delay ratio $\zeta$. While the output micro-pulse train has the smallest efficiency due to relatively large $\zeta$ value. As shown in Figs. 13 and 14, though the material dispersion of the shaper would influence the laser pulse passing through the shaper, the shaping algorithm can automatically adjust the shaper's elements rotation angles to compensate such influences and tailor the output pulse profile to match the desired shape.

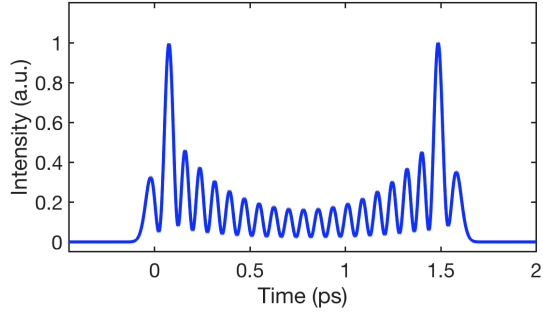

Fig. 12. Output pulse profile from the folded-type shaper ($B_1 = 1$, and $B_2 = 90°$) with initial elements rotation angles defined in Eq. (1). The input laser is unchirped Gaussian pulse with center wavelength of 266nm and duration of 80fs (FWHM). The phase delays of the BRs in the shaper are $0.1\pi$ rad, which is defined on basis of monochromatic wave with the center wavelength of the input pulse (i.e., 266nm). The shaper contains 20 a-cut $\alpha - BBO$ crystals as BRs with relative time delay ratio $\zeta = 0.62$.

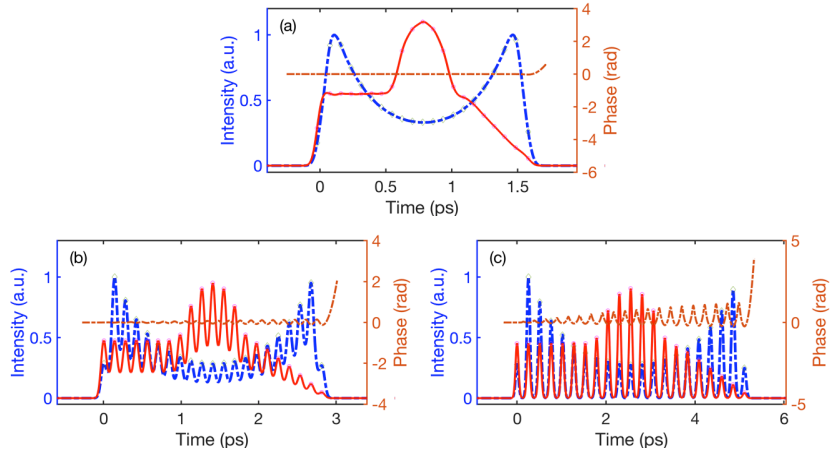

Fig. 13. Three different shaped pulses including (a) smooth profile, (b) rippled profile, (c) micro-pulse train, which are generated by the shaping algorithm in folded-type shaper ($B_1 = 1$, and $B_2 = 90°$) with material dispersion. The shaping efficiency for the three different pulses are about (1) 10.8%, (b) 6.3%, and (c) 5%. The blue dash-dot lines are the pulses from the shaper with initial elements rotation angles defined in Eq. (1). Its corresponding 21 reference points are shown with green diamond markers. The red solid lines are the shaped pulses generated by the shaping algorithm. Its corresponding 21 reference points are shown with magenta circle markers. The brown dash-dot lines are the phases of the shaped pulses in red solid lines. The input parameters $\{\delta, \sigma, \beta, \eta\}$ for the shaping algorithm are set to $\{1°, 1.4, 5, 0.2\%\}$. The input laser is 266nm unchirped femtosecond Gaussian pulse with duration of 80fs (FWHM). The shaper contains 20 BRs, which are $\alpha - BBO$ crystals of a-cut. The relative time delay ratios of the BRs are (a) $\zeta = 0.62$, (b) $\zeta = 1.1$, (c) $\zeta = 2$. The phase delays of the BRs are $\pi$ rad. For every figure that have phase lines in this paper, the shaped pulses phase (i.e. brown dash-dot lines) is shifted by a constant value, which is the difference between the phases (the time coordinate is the left end point of the brown dash-dot lines) of the initial output pulse (i.e. pulse in blue dash-dot line) and the shaped pulse (i.e. pulse in red sold line).

Table 10. Elements rotation angle variations $\Delta\Theta_n$ (Deg) for different types of shaped pulses showed in Fig. 13 (red solid lines) including (a) smooth profile, (b) rippled profile, (c) micro-pulse train, where $N = 20$, $n = 1, 2, \cdots, N, N + 1$ and $\Delta\Theta_{N+1} = \Delta\Theta_p$.

| Element number ($n$) | | Smooth | | Rippled | | Micro-pulse train | |
|---|---|---|---|---|---|---|---|
| 1 | 12 | -1.000 | -1.220 | -0.720 | -0.617 | -0.600 | -0.860 |
| 2 | 13 | -2.175 | -2.764 | -1.899 | -2.066 | -1.947 | -1.726 |
| 3 | 14 | -1.491 | -2.409 | -1.111 | -1.538 | -0.908 | -1.678 |
| 4 | 15 | -1.798 | -2.450 | -1.500 | -1.531 | -1.649 | -1.276 |
| 5 | 16 | -1.884 | -2.968 | -1.478 | -2.040 | -1.188 | -2.084 |
| 6 | 17 | -1.431 | -1.827 | -1.156 | -0.943 | -1.372 | -0.781 |
| 7 | 18 | -2.194 | -3.553 | -1.781 | -2.617 | -1.396 | -2.593 |
| 8 | 19 | -1.181 | -1.227 | -0.771 | -0.353 | -1.048 | -0.261 |
| 9 | 20 | -3.269 | -4.141 | -2.579 | -3.220 | -2.134 | -3.155 |
| 10 | 21 | -0.901 | -3.050 | -0.357 | -1.928 | -0.653 | -1.860 |
| 11 | — | -3.343 | — | -2.650 | — | -2.224 | — |

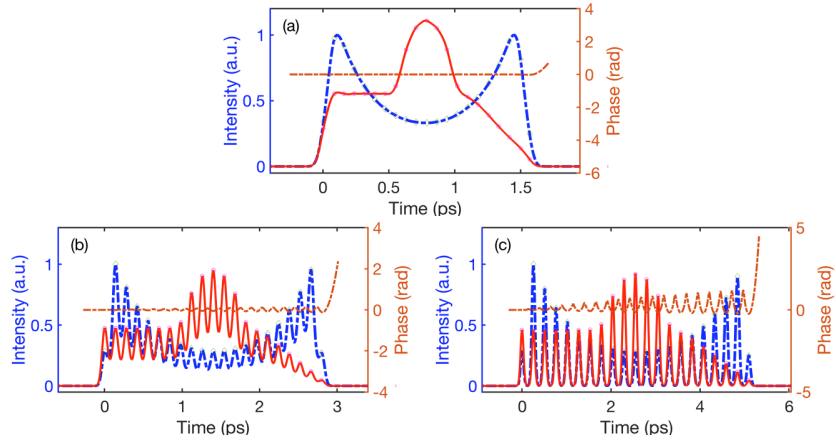

Fig. 14. Three different shaped pulses including (a) smooth profile, (b) rippled profile, (c) micro-pulse train, which are generated by the shaping algorithm in fan-type shaper ($B_1 = 1$, and $B_2 = 90°$) with material dispersion. The shaping efficiency for the three different pulses are about (a) 11%, (b) 6.3%, and (c) 5.1%. The blue dash-dot lines are the pulses from the shaper with initial elements rotation angles defined in Eq. (2). Its corresponding 21 reference points are shown with green diamond markers. The phase delays of the BRs are 0 rad. All other input parameters and meanings of the lines and markers are the same with those in Fig. 13 for folded-type shaper.

Table 11. Elements rotation angle variations $\Delta\Theta_n$ (Deg) for different types of shaped pulses showed in Fig. 14 (red solid lines) including (a) smooth profile, (b) rippled profile, (c) micro-pulse train, where $N = 20$, $n = 1, 2, \cdots, N$.

| BR number ($n$) | | Smooth | | Rippled | | Micro-pulse train | |
|---|---|---|---|---|---|---|---|
| 1 | 11 | 0.090 | 5.276 | 0.718 | 6.057 | 0.654 | 5.718 |
| 2 | 12 | -1.017 | 7.279 | -0.475 | 8.011 | -0.570 | 7.691 |
| 3 | 13 | -1.894 | 8.855 | -1.279 | 9.389 | -1.417 | 9.108 |
| 4 | 14 | -2.351 | 9.256 | -1.710 | 9.858 | -1.888 | 9.612 |
| 5 | 15 | -2.450 | 9.244 | -1.776 | 9.798 | -1.996 | 9.593 |
| 6 | 16 | -2.208 | 8.763 | -1.513 | 9.251 | -1.765 | 9.090 |
| 7 | 17 | -1.657 | 7.708 | -0.951 | 8.131 | -1.235 | 8.010 |
| 8 | 18 | -0.924 | 6.062 | -0.019 | 6.437 | -0.322 | 6.353 |
| 9 | 19 | 0.834 | 3.843 | 1.716 | 4.161 | 1.382 | 4.114 |
| 10 | 20 | 3.005 | 1.097 | 3.850 | 1.293 | 3.502 | 1.274 |

Due to wide bandwidth, femtosecond laser is easy to become chirped pulse when passing through the optical medium with material dispersion. Chirped laser pulses are very common in many practical situations. In this part, the shaping algorithm would be applied to arbitrary temporal shaping of chirped laser pulse. Figure 15 shows the temporal intensity profile and phase of a chirped laser pulse, which is generated by passing the unchirped Gaussian laser pulse through the quartz. The initial unchirped Gaussian pulse is 266nm laser with duration of 80fs (FWHM). Due to material dispersion, the laser pulse is broadened with duration being increased to ~100fs, as shown in Fig. 15. For chirped laser pulse shaping, the output pulse from the shaper can be calculated with the same method for unchirped femtosecond laser shaping as described above. Both the folded-type and fan-type shapers have 20 BRs, which are a-cut $\alpha - BBO$ crystals. The target pulse has three different types of temporal profiles including smooth, rippled and micro-pulse train type, which are listed in Table 4. Similar with the unchirped femtosecond laser pulse shaping above, the time coordinates of the $N + 1$ reference points need to be shifted because of the material dispersion. The new time coordinates of the $N + 1$ reference points can be determined with the same method as described above for unchirped femtosecond laser pulse shaping. The input parameters of the shaping algorithm both for folded-type and fan-type shapers are $\{\delta, \sigma, \beta, \eta\} = \{1°, 1.4, 5, 0.2\%\}$. With about $150 \sim 180$ iterations, the shaping algorithm automatically stops and outputs the results with $\eta_{out}$ less than the target value $\eta = 0.2\%$ for all three types of target pulses. The shaped pulses and the corresponding rotation angles adjustments are presented in Fig. 16, Table 12, Fig. 17 and Table 13. For chirped pulse shaping with shapers containing material dispersion, this algorithm can also automatically adjust the shaper's elements rotation angles to tailor the laser pulse toward desired profiles.

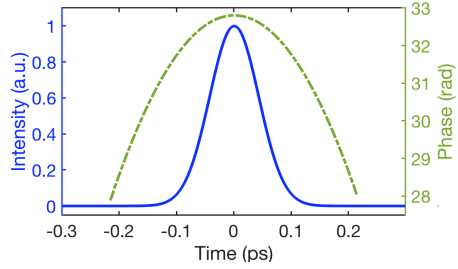

Fig. 15. Chirped 266nm laser pulse with duration of 100fs.

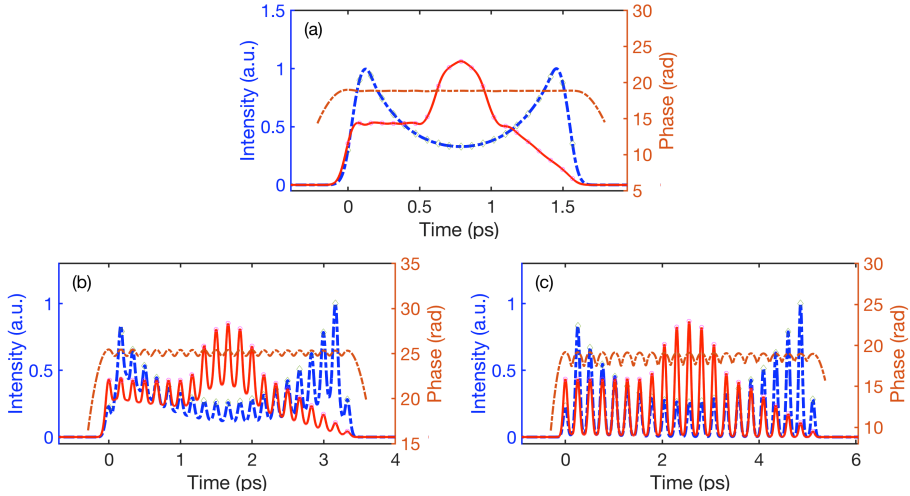

Fig. 16. Three different shaped pulses including (a) smooth profile, (b) rippled profile, (c) micro-pulse train, which are generated by the shaping algorithm in folded-type shaper ($B_1 = 1$, and $B_2 = 90°$ ) with material dispersion. The shaping efficiency for the three different pulses are about (a) 10.5%, (b) 5.6%, and (c) 5.1%. The input laser is 266nm chirped femtosecond pulse with duration of 100fs (FWHM) as shown in Fig. 15. The shaper contains 20 BRs, which are a-cut $\alpha - BBO$ crystals. The relative time delay ratios of the BRs are (a) $\zeta = 0.62$, (b) $\zeta = 1.3$, (c) $\zeta = 2$, which are defined on the basis of unchirped pulse duration (i.e., 80fs) of 266nm light. The phase delays of the BRs are $\pi$ rad. All other input parameters and meanings of the lines and markers are the same with those in Fig. 13 for folded-type shaper.

Table 12. Elements rotation angle variations $\Delta\Theta_n$ (Deg) for different types of shaped pulses showed in Fig. 16 (red solid lines) including (a) smooth profile, (b) rippled profile, (c) micro-pulse train, where $N = 20$, $n = 1, 2, \cdots, N, N + 1$ and $\Delta\Theta_{N+1} = \Delta\Theta_p$.

| Element number ($n$) | | Smooth | | Rippled | | Micro-pulse train | |
|---|---|---|---|---|---|---|---|
| 1 | 12 | -0.980 | -3.734 | -0.800 | -0.805 | -0.850 | -0.846 |
| 2 | 13 | -2.638 | -4.584 | -1.793 | -1.787 | -1.889 | -1.952 |
| 3 | 14 | -1.777 | -5.135 | -1.084 | -1.685 | -1.209 | -1.744 |
| 4 | 15 | -2.747 | -4.651 | -1.469 | -1.280 | -1.538 | -1.427 |
| 5 | 16 | -2.331 | -5.716 | -1.387 | -2.144 | -1.533 | -2.219 |
| 6 | 17 | -2.829 | -4.273 | -1.192 | -0.755 | -1.234 | -0.881 |
| 7 | 18 | -2.891 | -6.251 | -1.626 | -2.675 | -1.781 | -2.763 |
| 8 | 19 | -3.157 | -3.793 | -0.865 | -0.219 | -0.904 | -0.337 |
| 9 | 20 | -4.318 | -6.687 | -2.340 | -3.221 | -2.515 | -3.322 |
| 10 | 21 | -3.229 | -5.364 | -0.527 | -1.847 | -0.551 | -1.956 |
| 11 | — | -4.700 | — | -2.367 | — | -2.536 | — |

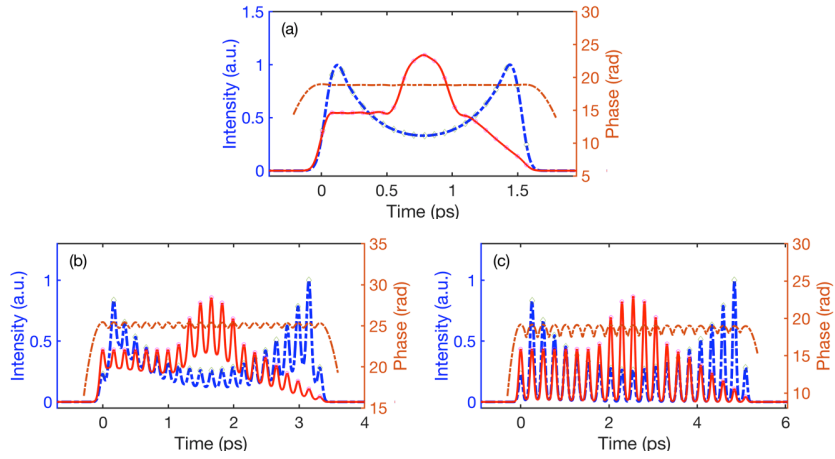

Fig. 17. Three different shaped pulses including (a) smooth profile, (b) rippled profile, (c) micro-pulse train, which are generated by the shaping algorithm in fan-type shaper ($B_1 = 1$, and $B_2 = 90°$) with material dispersion. The shaping efficiency for the three different pulses are about (a) 10.9%, (b) 5.7%, and (c) 5.1%. The blue dash-dot lines are the pulses from the shaper with initial elements rotation angles defined in Eq. (2). Its corresponding 21 reference points are shown with green diamond markers. The phase delays of the BRs are 0 rad. All other input parameters and meanings of the lines and markers are the same with those in Fig. 16 for folded-type shaper.

Table 13. Elements rotation angle variations $\Delta\Theta_n$ (Deg) for different types of shaped pulses showed in Fig. 17 (red solid lines) including (a) smooth profile, (b) rippled profile, (c) micro-pulse train, where $N = 20$, $n = 1, 2, \cdots, N$.

| BR number ($n$) | | Smooth | | Rippled | | Micro-pulse train | |
|---|---|---|---|---|---|---|---|
| 1 | 11 | 0.663 | 5.644 | 0.807 | 7.369 | 0.861 | 7.316 |
| 2 | 12 | -0.645 | 7.433 | -0.188 | 9.232 | -0.144 | 9.183 |
| 3 | 13 | -1.368 | 9.137 | -0.799 | 10.483 | -0.764 | 10.44 |
| 4 | 14 | -1.934 | 9.32 | -1.042 | 10.817 | -1.015 | 10.772 |
| 5 | 15 | -1.881 | 9.352 | -0.934 | 10.599 | -0.919 | 10.559 |
| 6 | 16 | -1.82 | 8.799 | -0.506 | 9.888 | -0.502 | 9.854 |
| 7 | 17 | -1.038 | 7.744 | 0.195 | 8.604 | 0.184 | 8.577 |
| 8 | 18 | -0.608 | 6.054 | 1.253 | 6.754 | 1.216 | 6.733 |
| 9 | 19 | 1.329 | 3.848 | 3.043 | 4.346 | 3.002 | 4.334 |
| 10 | 20 | 3.283 | 1.042 | 5.199 | 1.366 | 5.147 | 1.362 |

The Appendix in this paper presents the simulation results for chirped and unchirped femtosecond laser shaping in shapers with nonidentical BRs time delays and phase delays, as shown in Figs. (A3-A6) and Tables A1-A4. The shaping algorithm automatically stops and outputs the results after about 150 ~ 180 iterations with shaping errors $\eta_{out}$ less than the target error $\eta = 0.2\%$. Though the deviations introduced to the BRs time delays and phase delays would influence the shaped pulse profile, the shaping algorithm can automatically adjust shaper's elements rotation angles to make the shaped pulse profile approach to the desired shape. The shaping accuracies $\eta_{out}$ for the $N+1$ reference points still can be reduced to smaller than the target value $\eta$. When these deviations aren't very big, the folded-type and fan-type shapers can produce shaped pulses which are very close to the target pulses. For the overall profile of the output pulse of the shaper, it can be improved by reducing the deviations introduced to the BRs of the shaper.

## 6. Discussion

The shaping algorithm in this paper doesn't require the precise knowledge about the shaper and the input pulse. Therefore, this shaping method has relatively strong robustness for such nonlinear optimization problems of pulse shaping. By discretizing the target optical profile with $N+1$ points, the complex target profile is able to be connected with the elements of the shaper containing $N$ number of BRs. This shaping method may be applicable to non-ideal shapers and input lasers, such as nonidentical BRs time delays and phase delays, material dispersion, and chirped laser, etc. For some applications in which the input laser pulse, desired output pulse of the shaper or shaper's characteristics can't be determined exactly beforehand, this shaping algorithm may also be applicable to perform the shaping task. These situations are very common in many practical applications. The shaping method introduced in this paper can help to generate high fidelity shaped pulse with various target profiles. For folded-type or fan-type shapers that contain large number of BRs, this shaping algorithm would be a great advantage, which can automatically adjust the shaper to generate the desired pulse rapidly. Compared to the traditional optimization algorithms (e.g. simulated annealing algorithms, particle swarm optimization algorithms, genetic algorithms, ···), which may also be applicable for such nonlinear optimization problems, the shaping algorithm in this paper requires lesser number of iterations and has high accuracy of convergence. With the properties of folded-type and fan-type shapers' components, this shaping method may be applicable to laser shaping over a wide range of laser parameters (e.g. laser wavelength from infrared to ultraviolet, laser pulse duration from femtosecond to picosecond, arbitrary pulse repetition rate, high power, etc.). The output shaped pulses of the folded-type and fan-type shapers are linearly polarized, and their energy can be amplified by laser power amplifiers. This shaping method may also be applied to adaptive shaping in fields like nonlinear frequency conversion, laser power amplifiers, electron bunch stimulated from photocathode, and so on. For these applications, the temporal profiles of the output optical pulse or electron bunch may be different from that of input drive lasers due to the nonlinear processes. Besides the output pulse of the shaper, the processes (e.g. second harmonic pulses, electron bunches) induced by the input drive laser may also be used as feedback signal for this shaping algorithm to directly optimize the shaper to perform adaptive shaping (e.g. generating second harmonic pulse or electron bunch with various target temporal profiles, and so on).

## 7. Summary

In summary, a new shaping algorithm is proposed and applicable to folded-type and fan-type shapers for generation of arbitrary optical waveforms. This shaping algorithm can be applied to shapers containing large number of BRs and be able to automatically generate shaped pulses of desired profiles with high fidelity. The shaping method in this paper has strong robustness and may be applicable to non-ideal shapers or input lasers (e.g. with nonidentical BRs, material dispersion, and chirped laser, etc.). For some situations where the complete knowledge for pulse shaping (e.g. input pulse, desired output pulse of the shaper or the shaper's characteristics) can't be determined exactly beforehand, this shaping method may also be applicable to perform the shaping task. This research may bring new opportunities for many potential applications over a wide range of laser parameters.

## 8. Appendix

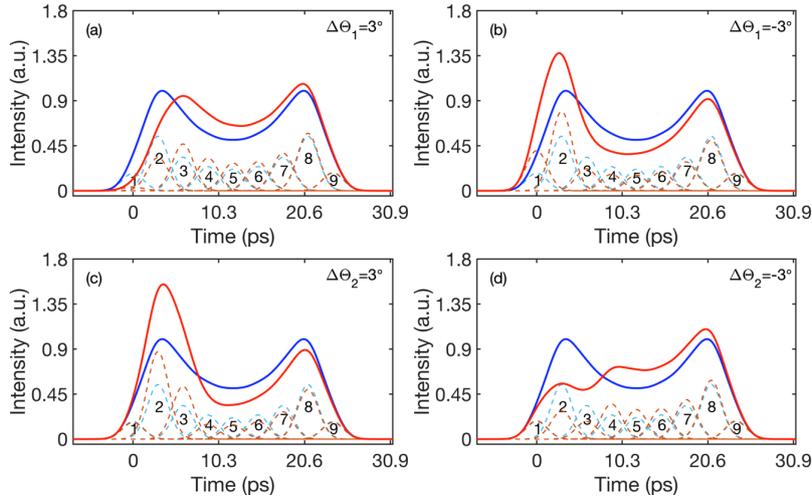

Fig. A1. Variations of the replicas and shaped pulses when changing the rotation angles of the 1st and 2nd BRs in the folded-type shaper with $B_1 = 1$ and $B_2 = 90°$. The blue solid lines are the temporal intensity profiles of shaped pulses from the shaper with initial elements rotation angles as defined in Eq. (1). Its corresponding $N + 1$ replicas are represented with cyan dashed lines. The red solid lines are the temporal intensity profiles of the shaped pulses from the shaper with only $\Theta_{1, or\, 2}$ changed by $3°$ or $-3°$, and all other input parameters are the same with those for the pulse in blue solid line. Its corresponding $N + 1$ replicas are represented with brown dashed lines. The input laser Gaussian pulse with 3ps (FWHM) duration. The shaper has eight BRs with relative time delay ratios $\zeta$ of 1 and phase delays of $\pi$ rad.

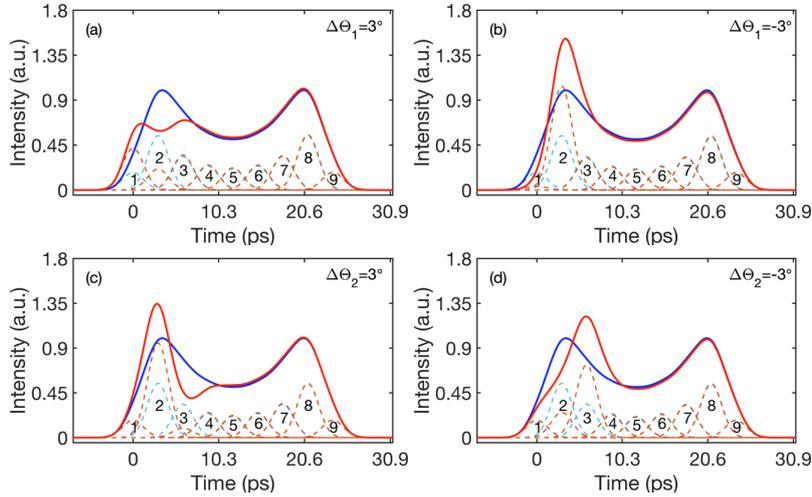

Fig. A2. Variations of the replicas and shaped pulses when changing the rotation angles of the 1st and 2nd BRs in the fan-type shaper. The blue solid lines are the temporal intensity profiles of shaped pulses from the shaper with initial elements rotation angles as defined in Eq. (2). The corresponding $N + 1$ replicas are represented with cyan dashed lines. The BRs phase delays are set to 0 rad. All other input parameters and meanings of lines are the same with those in Fig. A1.

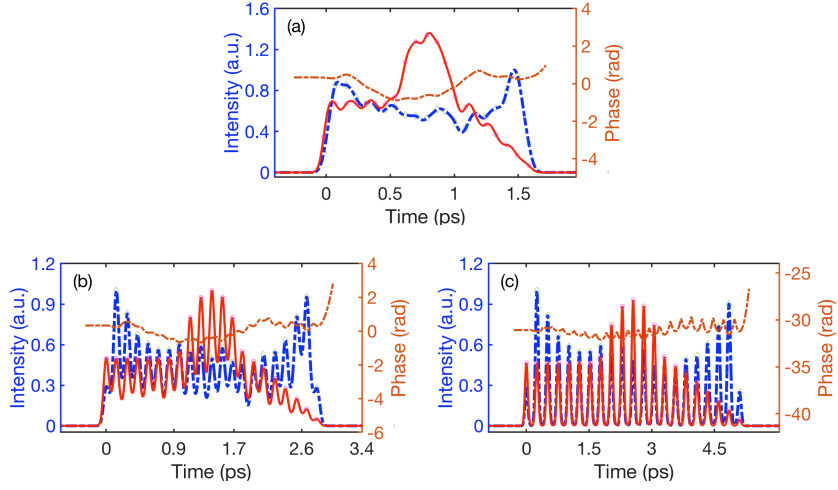

Fig. A3. Three different shaped pulses including (a) smooth profile, (b) rippled profile, (c) micro-pulse train, which are generated by the shaping algorithm in folded-type shaper ($B_1 = 1$, and $B_2 = 90°$) with material dispersion. The shaping efficiency for the three different pulses are about (1) 13.9%, (b) 6.6%, and (c) 5.2%. The relative intensities of the 21 reference points for three different types of target pulses are listed in Table 4. However, the relative intensity of the 1st reference point for target pulse of smooth profile is 0.52. The blue dash-dot lines are the pulses from the shaper with initial elements rotation angles defined in Eq. (1). Its corresponding 21 reference points are shown with green diamond markers. The red solid lines are the shaped pulses generated by the shaping algorithm. Its corresponding 21 reference points are shown with magenta circle markers. The brown dash-dot lines are the phases of the shaped pulses in red solid lines. The input parameters $\{\delta, \sigma, \beta, \eta\}$ for the shaping algorithm are set to $\{1°, 1.4, 5, 0.2\%\}$. The input laser is 266nm unchirped femtosecond Gaussian pulse with duration of 80fs (FWHM). The shaper contains 20 BRs, which are $\alpha - BBO$ crystals of a-cut. The BRs of the shaper are introduced with deviations as shown in Eq. (9), Eq. (10) and Table 6. The tolerances $\{W_1, W_2\}$ for target pulses of (a) smooth profile, (b) rippled profile, and (c) micro-pulse train are $\{4fs, 0.1\pi\}$, $\{4fs, 0.1\pi\}$, and $\{4fs, 0.1\pi\}$, respectively. The nominal relative time delay ratios of the BRs are (a) $\zeta = 0.62$, (b) $\zeta = 1.1$, (c) $\zeta = 2$. The nominal phase delays of the BRs are $\pi$ rad. For every figure that have phase lines in this paper, the shaped pulses phase (i.e. brown dash-dot lines) is shifted by a constant value, which is the difference between the phases (the time coordinate is the left end point of the brown dash-dot lines) of the initial output pulse (i.e. pulse in blue dash-dot line) and the shaped pulse (i.e. pulse in red sold line).

Table A1. Elements rotation angle variations $\Delta\Theta_n$ (Deg) for different types of shaped pulses showed in Fig. A3 (red solid lines) including (a) smooth profile, (b) rippled profile, (c) micro-pulse train, where $N = 20$, $n = 1, 2, \cdots, N, N+1$ and $\Delta\Theta_{N+1} = \Delta\Theta_p$.

| Element number ($n$) | | Smooth | | Rippled | | Micro-pulse train | |
|---|---|---|---|---|---|---|---|
| 1 | 12 | -1.300 | -2.225 | -0.720 | -1.601 | -0.600 | -1.600 |
| 2 | 13 | -1.816 | -4.282 | -2.051 | -1.792 | -2.035 | -1.686 |
| 3 | 14 | -2.021 | -2.749 | -0.937 | -2.270 | -0.794 | -2.280 |
| 4 | 15 | -1.269 | -4.280 | -1.840 | -1.515 | -1.845 | -1.408 |
| 5 | 16 | -2.168 | -4.342 | -1.044 | -2.642 | -0.897 | -2.599 |
| 6 | 17 | -1.133 | -4.443 | -1.753 | -1.175 | -1.753 | -1.052 |
| 7 | 18 | -2.278 | -5.536 | -1.062 | -3.081 | -0.911 | -3.019 |
| 8 | 19 | -1.363 | -3.843 | -1.643 | -0.675 | -1.651 | -0.594 |
| 9 | 20 | -4.135 | -6.596 | -1.842 | -3.580 | -1.659 | -3.514 |
| 10 | 21 | -1.215 | -5.619 | -1.343 | -2.277 | -1.365 | -2.223 |
| 11 | — | -4.657 | — | -2.068 | — | -1.908 | — |

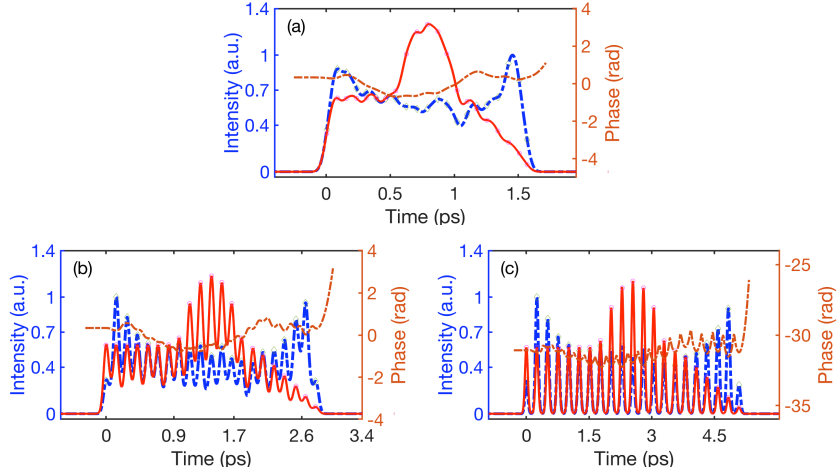

Fig. A4. Three different shaped pulses including (a) smooth profile, (b) rippled profile, (c) micro-pulse train, which are generated by the shaping algorithm in fan-type shaper ($B_1 = 1$, and $B_2 = 90°$) with material dispersion. The shaping efficiency for the three different pulses are about (1) 13.2%, (b) 7.8%, and (c) 6.3%. The blue dash-dot lines are the pulses from the shaper with initial elements rotation angles defined in Eq. (2). Its corresponding 21 reference points are shown with green diamond markers. The nominal phase delays of the BRs are 0 rad. All other input parameters and meanings of the lines and markers are the same with those in Fig. A3 for folded-type shaper.

Table A2. Elements rotation angle variations $\Delta\Theta_n$ (Deg) for different types of shaped pulses showed in Fig. A4 (red solid lines) including (a) smooth profile, (b) rippled profile, (c) micro-pulse train, where $N = 20$, $n = 1, 2, \cdots, N$.

| BR number ($n$) | | Smooth | | Rippled | | Micro-pulse train | |
|---|---|---|---|---|---|---|---|
| 1 | 11 | 0.103 | 4.399 | 1.045 | 5.115 | 0.976 | 4.737 |
| 2 | 12 | -0.570 | 5.905 | 0.194 | 6.634 | 0.089 | 6.249 |
| 3 | 13 | -1.136 | 6.982 | -0.295 | 7.932 | -0.435 | 7.627 |
| 4 | 14 | -1.228 | 7.791 | -0.417 | 8.451 | -0.595 | 8.190 |
| 5 | 15 | -1.326 | 8.008 | -0.367 | 8.625 | -0.566 | 8.384 |
| 6 | 16 | -1.212 | 7.619 | -0.162 | 8.229 | -0.386 | 8.064 |
| 7 | 17 | -1.070 | 6.900 | 0.065 | 7.395 | -0.194 | 7.247 |
| 8 | 18 | -0.988 | 5.740 | 0.359 | 5.968 | 0.063 | 5.849 |
| 9 | 19 | 0.445 | 3.653 | 1.622 | 3.895 | 1.287 | 3.833 |
| 10 | 20 | 2.327 | 1.105 | 3.184 | 1.200 | 2.808 | 1.167 |

Fangming Liu 26

Fig. A5. Three different shaped pulses including (a) smooth profile, (b) rippled profile, (c) micro-pulse train, which are generated by the shaping algorithm in folded-type shaper ($B_1 = 1$, and $B_2 = 90°$) with material dispersion. The relative intensities of the 21 reference points for three different types of target pulses are listed in Table 4. The shaping efficiency for the three different pulses are about (1) 9.7%, (b) 5.9%, and (c) 5.4%. The blue dash-dot lines are the pulses from the shaper with initial elements rotation angles defined in Eq. (1). Its corresponding 21 reference points are shown with green diamond markers. The red solid lines are the shaped pulses generated by the shaping algorithm. Its corresponding 21 reference points are shown with magenta circle markers. The brown dash-dot lines are the phases of the shaped pulses in red solid lines. The input parameters $\{\delta, \sigma, \beta, \eta\}$ for the shaping algorithm are set to $\{1°, 1.4, 5, 0.2\%\}$. The input laser is 266nm chirped femtosecond pulse with duration of 100fs (FWHM) as shown in Fig. 15. The shaper contains 20 BRs, which are $\alpha - BBO$ crystals of a-cut. The BRs of the shaper are introduced with deviations as shown in Eq. (9), Eq. (10) and Table 6. The tolerances $\{W_1, W_2\}$ for target pulses of (a) smooth profile, (b) rippled profile, and (c) micro-pulse train are $\{4fs, 0.02\pi\}$, $\{4fs, 0.1\pi\}$, and $\{4fs, 0.1\pi\}$, respectively. The nominal relative time delay ratios of the BRs are (a) $\zeta = 0.62$, (b) $\zeta = 1.3$, (c) $\zeta = 2$. The nominal phase delays of the BRs are $\pi$ rad.

Table A3. Elements rotation angle variations $\Delta\Theta_n$ (Deg) for different types of shaped pulses showed in Fig. A5 (red solid lines) including (a) smooth profile, (b) rippled profile, (c) micro-pulse train, where $N = 20$, $n = 1, 2, \cdots, N, N + 1$ and $\Delta\Theta_{N+1} = \Delta\Theta_p$.

| Element number ($n$) | | Smooth | | Rippled | | Micro-pulse train | |
|---|---|---|---|---|---|---|---|
| 1 | 12 | -1.000 | -6.126 | -0.800 | -1.720 | -0.850 | -1.726 |
| 2 | 13 | -3.100 | -6.160 | -1.975 | -1.814 | -2.023 | -1.814 |
| 3 | 14 | -2.194 | -7.065 | -1.004 | -2.428 | -1.046 | -2.426 |
| 4 | 15 | -3.613 | -6.160 | -1.778 | -1.486 | -1.831 | -1.504 |
| 5 | 16 | -2.729 | -7.645 | -1.092 | -2.769 | -1.137 | -2.781 |
| 6 | 17 | -3.900 | -6.060 | -1.710 | -1.099 | -1.758 | -1.136 |
| 7 | 18 | -3.474 | -8.230 | -1.085 | -3.189 | -1.130 | -3.212 |
| 8 | 19 | -4.609 | -5.718 | -1.628 | -0.646 | -1.672 | -0.680 |
| 9 | 20 | -5.319 | -8.928 | -1.809 | -3.684 | -1.853 | -3.703 |
| 10 | 21 | -4.877 | -7.530 | -1.398 | -2.317 | -1.421 | -2.341 |
| 11 | — | -5.994 | — | -2.044 | — | -2.072 | — |

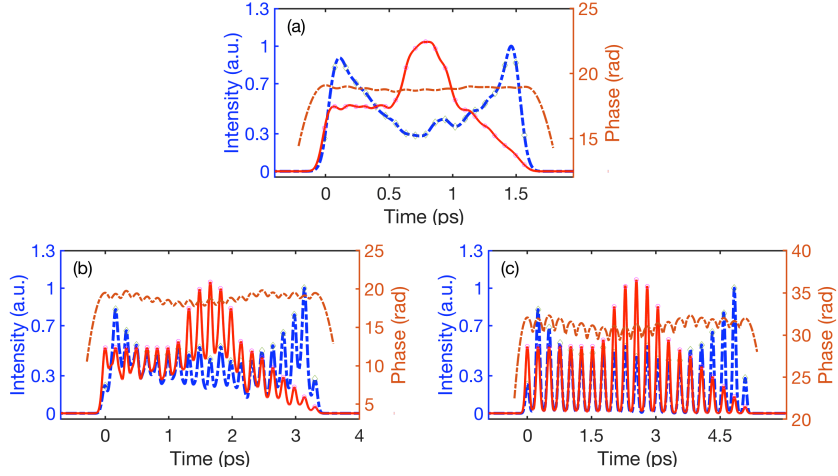

Fig. A6. Three different shaped pulses including (a) smooth profile, (b) rippled profile, (c) micro-pulse train, which are generated by the shaping algorithm in fan-type shaper ($B_1 = 1$, and $B_2 = 90°$) introduced with material dispersion and deviations for BRs time delays and phase delays. The shaping efficiency for the three different pulses are about (1) 11%, (b) 7%, and (c) 6.4%. The blue dash-dot lines are the pulses from the shaper with initial elements rotation angles defined in Eq. (2). Its corresponding 21 reference points are shown with green diamond markers. The nominal phase delays of the BRs are 0 rad. All other input parameters and meanings of the lines and markers are the same with those in Fig. A5 for folded-type shaper.

Table A4. Elements rotation angle variations $\Delta\Theta_n$ (Deg) for different types of shaped pulses showed in Fig. A6 (red solid lines) including (a) smooth profile, (b) rippled profile, (c) micro-pulse train, where $N = 20$, $n = 1, 2, \cdots, N$.

| BR number ($n$) | | Smooth | | Rippled | | Micro-pulse train | |
|---|---|---|---|---|---|---|---|
| 1 | 11 | 1.290 | 6.459 | 1.153 | 6.467 | 1.201 | 6.382 |
| 2 | 12 | -0.495 | 7.685 | 0.507 | 7.870 | 0.541 | 7.804 |
| 3 | 13 | -0.711 | 9.000 | 0.223 | 9.113 | 0.236 | 9.029 |
| 4 | 14 | -1.495 | 9.414 | 0.295 | 9.511 | 0.295 | 9.422 |
| 5 | 15 | -1.313 | 9.423 | 0.546 | 9.498 | 0.530 | 9.423 |
| 6 | 16 | -1.632 | 8.838 | 0.931 | 8.968 | 0.900 | 8.901 |
| 7 | 17 | -0.611 | 7.879 | 1.303 | 7.923 | 1.248 | 7.871 |
| 8 | 18 | -0.656 | 6.281 | 1.696 | 6.303 | 1.634 | 6.269 |
| 9 | 19 | 1.697 | 4.117 | 3.025 | 4.092 | 2.942 | 4.075 |
| 10 | 20 | 3.620 | 1.122 | 4.547 | 1.265 | 4.488 | 1.263 |